\documentclass[UTF8,12pt,journal,draftclsnofoot,oneside,onecolumn]{IEEEtran}

\usepackage[cmex10]{amsmath}
\usepackage{amssymb}
\usepackage{epsfig}
\usepackage{subfigure}
\usepackage{cite}
\usepackage{graphicx}
\usepackage{color}
\usepackage{bm}
\usepackage{booktabs}
\usepackage{gensymb}
\newlength{\figwidth}
\newcommand{\tabincell}[2]{\begin{tabular}{@{}#1@{}}#2\end{tabular}}

\newtheorem{lemma}{\it Lemma}

\newtheorem{remark}{\it Remark}
\newtheorem{proposition}{\it Proposition}
\makeatletter

\begin{document}

\title{On the Secrecy Design of STAR-RIS assisted Uplink NOMA Networks}
\author{Zheng Zhang, Jian Chen,~\IEEEmembership{Member,~IEEE}, Yuanwei Liu,~\IEEEmembership{Senior Member,~IEEE}, \\ Qingqing Wu,~\IEEEmembership{Member,~IEEE}, Bingtao He, and Long Yang,~\IEEEmembership{Member,~IEEE} \vspace{-10.5mm}
\thanks{Zheng Zhang, Jian Chen, Bingtao He, and Long Yang are with the State Key Laboratory of Integrated Services Networks, Xidian University, Xi'an 710071, China (e-mail: zzhang\_688@stu.xidian.edu.cn; jianchen@mail.xidian.edu.cn; bthe@xidian.edu.cn; lyang@xidian.edu.cn;). Yuanwei Liu is with the School of Electronic Engineering and Computer Science, Queen Mary University of London, London E1 4NS, U.K. (e-mail: yuanwei.liu@qmul.ac.uk). Qingqing Wu is with the State Key Laboratory of Internet of Things for Smart City, University of Macau, Macau 999078, China (e-mail: qingqingwu@um.edu.mo).}
}
\maketitle
\begin{abstract}
    This paper investigates the secure transmission in a simultaneously transmitting and reflecting reconfigurable intelligent surface (STAR-RIS) assisted uplink non-orthogonal multiple access system, where the legitimate users send confidential signals to the base station by exploiting STAR-RIS to reconfigure the electromagnetic propagation environment proactively. Depending on the availability of the eavesdropping channel state information (CSI), both the full CSI and statistical CSI of the eavesdropper are considered. For the full eavesdropping CSI scenario, we adopt the adaptive-rate wiretap code scheme with the aim of maximizing minimum secrecy capacity subject to the successive interference cancellation decoding order constraints. To proceed, we propose an alternating hybrid beamforming (AHB) algorithm to jointly optimize the receive beamforming, transmit power, and reflection/transmission coefficients. While for the statistical eavesdropping CSI scenario, the constant-rate wiretap code scheme is employed to minimize the maximum secrecy outage probability (SOP) subject to the quality-of-service requirements of legitimate users. Then, we derive the exact SOP expression under the constant-rate coding strategy and develop an extended AHB algorithm for the joint secrecy beamforming design. Simulation results demonstrate the effectiveness of the proposed scheme. Moreover, some useful guidance about the quantification of phase shift/amplitude and the deployment of STAR-RIS is provided.
\end{abstract}

\IEEEpeerreviewmaketitle

\section{Introduction}

Driven by the growing demands for high-data-rate required applications and ubiquitous wireless connectivity, reconfigurable intelligent surface (RIS) is emerged as a promising technique for next-generation wireless communications \cite{M.Di.Renzo_RIS_JSAC,R.Zhang_IRS_magazine}. Generally, RIS is a type of metamaterial-based planar array, consisting of plenty of passive scattering elements. Each reconfigurable element is able to adjust the phase shift and amplitude independently via a smart controller, thus controlling the radio propagation environment proactively. However, due to the hardware implementation, conventional RISs can only reflect the incident signals and serve wireless devices located on the same side of it, which immensely limits their deployment flexibility and coverage ranges \cite{R.Zhang_IRS_passive,R.Zhang_IRS_discrete,Wu_IRS_tutorial}. To address this, a novel type of metamaterial, namely simultaneous transmitting and reflecting RIS (STAR-RIS) \cite{Liu_magazine,X.Mu_STAR-RIS,J.Xu_STAR-RIS}, is proposed, which supports both electric-polarization and magnetization currents \cite{NTT_DOCOMO,C.Pfeiffer_STAR-RIS}, so as to reflect and/or transmit the incident signals upon it. Compared with the conventional RIS, STAR-RIS is able to provide a full-space (i.e., 360$^{\circ}$) service coverage and enjoys a more flexible deployment.

As another enabling technique, non-orthogonal multiple access (NOMA) can serve multiple users within one spectrum resource by leveraging the successive interference cancellation (SIC), to achieve massive connectivity and spectral efficiency enhancement \cite{L.Lv_magazine}. Compared with orthogonal multiple access (OMA) transmission, NOMA is more sensitive to the wireless channels as it introduces additional co-channel interference. The integration of NOMA and STAR-RIS can naturally overcome the randomness of wireless channels, thus further boosting the spectral performance  \cite{C.Wu_STAR-RIS_NOMA,W.Ni_STAR-RIS_NOMA,J.Zuo_STAR-RIS_NOMA,T.Hou_STAR-RIS_NOMA,Z.Xie_STAR-RIS_NOMA}. The authors of \cite{C.Wu_STAR-RIS_NOMA} investigated the fundamental coverage characterization of the STAR-RIS assisted NOMA networks, which demonstrated the superior coverage performance of STAR-RIS than conventional RIS. The work of \cite{W.Ni_STAR-RIS_NOMA} proposed a novel NOMA integrated over-the-air federated learning (AirFL) framework, where STAR-RIS was deployed to enhance spectral efficiency and mitigate interference. In \cite{J.Zuo_STAR-RIS_NOMA}, the authors studied the resource allocation problem of the STAR-RIS assisted NOMA network, where a two-layer iterative algorithm was developed to jointly optimize transmit beamforming and reflection/transmission beamforming. Furthermore, the STAR-RIS assisted NOMA enhanced coordinated multi-point transmission (CoMP) transmission was investigated in \cite{T.Hou_STAR-RIS_NOMA}. In \cite{Z.Xie_STAR-RIS_NOMA}, the authors further proposed a general analytical framework for STAR-RIS assisted NOMA networks.

Nevertheless, due to the broadcast characteristic of wireless transmissions, confidential or sensitive data is exposed to vulnerable communication environments. For the provision of the secure and private information transmission, the concept of physical layer security (PLS) is proposed from the perspective of information theory, which exploits the inherent characteristics of wireless channels (e.g., noise, fading, and interference) to degrade the legitimate information leakage \cite{Wyner,Y.Liu_NOMA_PLS}. Benefiting from the RIS's ability to adjust wireless channels, the integration of RIS into the PLS has received much attention \cite{M.Cui_IRS_PLS,X.Guan_IRS_AN,L.Dong_Without_CSI,ZZ_Robust,ZZ_statistical}. Specifically, The authors of \cite{M.Cui_IRS_PLS} studied the secrecy capacity maximization problem of the RIS assisted multiple-input-single-output (MISO) network, which validated the potential of integrating RIS into PLS. A novel artificial noise (AN) based jamming protocol via RIS was proposed in \cite{X.Guan_IRS_AN}, which shows that AN was conducive to the secrecy enhancement of the RIS assisted networks. The work of \cite{L.Dong_Without_CSI} further proposed a robust AN aided transmission scheme, which guarantees the secrecy performance without requiring eavesdropping channel state information (CSI). While for NOMA networks, the authors of \cite{ZZ_Robust} and \cite{ZZ_statistical} studied the secure transmission problems of the RIS assisted NOMA networks under the imperfect and statistical eavesdropping CSI, respectively. More recently, an initial exploration on STAR-RIS assisted PLS was studied in \cite{H.Niu_STAR-RIS_security}, which showed that compared with conventional RIS, STAR-RIS is capable of providing higher secrecy performance.

\vspace{-4mm}
\subsection{Motivations and Contributions}
\vspace{-2mm}
From the aforementioned efforts \cite{C.Wu_STAR-RIS_NOMA,W.Ni_STAR-RIS_NOMA,J.Zuo_STAR-RIS_NOMA,T.Hou_STAR-RIS_NOMA,Z.Xie_STAR-RIS_NOMA,M.Cui_IRS_PLS,X.Guan_IRS_AN,L.Dong_Without_CSI,ZZ_Robust,ZZ_statistical,H.Niu_STAR-RIS_security}, two significant observations are obtained as follows.
\begin{itemize}
  \item Although many efforts \cite{C.Wu_STAR-RIS_NOMA,W.Ni_STAR-RIS_NOMA,J.Zuo_STAR-RIS_NOMA,T.Hou_STAR-RIS_NOMA,Z.Xie_STAR-RIS_NOMA} are devoted to the STAR-RIS assisted NOMA communications, the critical issue of securing privacy has not been addressed yet, to the best of our knowledge. In fact, since STAR-RIS is capable of providing a 360$^{\circ}$ service, it inevitably leads to a 360$^{\circ}$ eavesdropping, which brings more serious security hazards to the private information transmission than that with conventional reflecting-only RIS.
  \item Meanwhile, the existing schemes devoted to the RIS/STAR-RIS enhanced PLS \cite{M.Cui_IRS_PLS,X.Guan_IRS_AN,L.Dong_Without_CSI,ZZ_Robust,ZZ_statistical,H.Niu_STAR-RIS_security} are not applicable to the STAR-RIS assisted uplink transmission due to its unique energy splitting (ES) protocol. Specifically, since both the incident signals from the different sides of STAR-RIS suffer reflection and transmission propagations, a new type of energy/signal leakage, namely \textit{opposite-side leakage}, appears, i.e., the reflecting/transmitting signals upon the opposite side of STAR-RIS from the base station (BS) are fully leaked, which requires the novel secrecy beamforming framework.
\end{itemize}

Motivated by the above, we focus on a STAR-RIS assisted uplink NOMA secrecy communication, where both full eavesdropping CSI and statistical eavesdropping CSI are considered. Our goal is to achieve the active control of the wireless propagation environment via the STAR-RIS, thus protecting the NOMA communication from malicious eavesdropping. The main contributions of this paper are summarized as follows.
\begin{itemize}
  \item We propose a STAR-RIS assisted uplink NOMA communication framework in the presence of an eavesdropper, where the indoor user (IU) and outdoor user (OU) send their signals to the BS through the reflection/transmission links provided by STAR-RIS. Depending on the availability of instantaneous eavesdropping CSI, two types of eavesdropping CSI are considered, i.e., full eavesdropping CSI and statistical eavesdropping CSI. Accordingly, we respectively formulate the joint beamforming and power optimization problems for minimum secrecy capacity maximization and maximum SOP minimization, subject to the SIC decoding constraints and reflection/transmission coefficient constraints.
  \item To tackle the non-convex optimization problem under the full eavesdropping CSI, we decompose the original problem into two subproblems and develop an alternating hybrid beamforming (AHB) algorithm for secrecy beamforming design. For the joint beamforming optimization subproblem, a polarization identity based two-layer iterative algorithm is designed to obtain the stationary point solution of the receive/passive beamforming. While for the transmit power optimization subproblem, the optimal transmit power policy is derived in the closed-form expression. Moreover, the convergence of the proposed AHB algorithm is theoretically proved.
  \item To handle the more challenging problem for the case of statistical eavesdropping CSI, we first derive the exact SOP expression, which converts the non-convex objective function to a tractable form. Then, an extended AHB algorithm is proposed to convexify the non-convex constraints, thus minimizing the maximum SOP among legitimate users, where we modify the two-layer iterative algorithm to optimize the hybrid beamforming, and derive the closed-form optimal transmit power strategy based on the monotonicity analysis.
  \item Simulation results demonstrate the convergence of the proposed algorithms and draw three interesting insights. 1) \textit{\textbf{Secrecy superiority}:} STAR-RIS assisted NOMA communication achieves the best secrecy performance over other baseline schemes. 2) \textit{\textbf{Secrecy quantization loss}:} when neglecting the existence of eavesdropper, the transmission rate only requires the $3$-bit quantization to achieve $98.07\%$ performance of the continuous phase shifts/amplitudes, which is consistent with the conventional reflecting-only RIS. However, when considering the secrecy transmission against the eavesdropper, the secrecy rate needs $4$ or more bits for achieving around $98\%$ performance of the continuous case. 3) \textit{\textbf{Secrecy deployment guidance}:} deploying STAR-RIS near the users or the BS under the adaptive-rate wiretap code setting is preferred, whereas it is preferable to deploy STAR-RIS far away from both of them under the constant-rate wiretap code setting.
\end{itemize}

The remainder of this paper is as follows. In Section II, we present the system model and formulate the optimization problems. Section III develops an AHB algorithm for the challenging optimization problem with the full eavesdropping CSI case. Section IV extends the AHB algorithm proposed in the previous section to the case of statistical eavesdropping CSI. Our numerical results and discussions are shown in Section V. Finally, the conclusions are presented in Section VI.

\textit{Notations:} boldface capital $\mathbf{X}$ and lower-case letter $\mathbf{x}$ represent matrix and vector respectively. $\mathbf{X}\in\mathbb{C}^{N\times M}$ denotes the complex-valued matrix lying in the $N\times M$-dimensional space, while $\mathbf{X}^{T}$ and $\mathbf{X}^{H}$ are the transpose and Hermitian conjugate operations. The rank value, trace operation and spectral norm of X are respectively denoted by $\text{rank}(\mathbf{X})$, $\text{Tr}(\mathbf{X})$ and $\|\mathbf{X}\|_{2}$. The positive semidefinite matrix is denoted by $\mathbf{X}\succeq\mathbf{0}$, while the circularly symmetric complex Gaussian
(CSCG) vector with zero mean and covariance matrix $\mathbf{X}$ is $\mathbf{x}\sim \mathcal{CN}(0,\mathbf{X})$. The diagonal matrix
whose main diagonal elements equal to the elements of vector x is denoted by $\text{diag}(\mathbf{x})$, while $\sigma_{i}$
denotes the \textit{i}th largest singular value of corresponding matrix. $|x|$ and $\|\mathbf{x}\|$ respectively denote
the modulus of complex variable $x$ and the Euclidean norm of the vector $\mathbf{x}$. $\mathbb{P}($\textperiodcentered$)$ and $\mathbb{E}($\textperiodcentered$)$ are the probability operation and statistical expectation. $\mathbf{I}$ is the identity matrix, $\Re(\cdot)$ denotes the real component of the complex value, while $\odot$ denotes the Hadamard product.

\vspace{-2mm}
\section{System Model and Problem Formulation}\vspace{-1mm}
\begin{figure}[t]
  \centering
  \includegraphics[scale = 0.37]{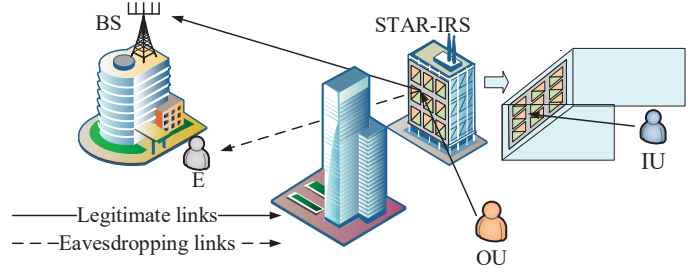}
  \caption{A STAR-RIS assisted uplink NOMA secrecy communication system.} \vspace{-8mm}
  \label{Fig.1}
\end{figure}
We consider a STAR-RIS assisted uplink NOMA network as shown in Fig. \ref{Fig.1}, which consists of an IU, an OU, a BS, an eavesdropper (E), and a STAR-RIS. E is located at the BS end and tries to wiretap both signals from IU and OU. There is no direct link between the BS/E and IU/OU due to the obstacles. To resist eavesdropping while relaying the superimposed signals from IU/OU to the BS, we deploy a STAR-RIS on the building facades, with being operating in the ES mode \cite{Liu_magazine}. All the nodes are assumed to operate in a half-duplex mode. The BS is equipped with $M$ antennas, while the remaining nodes are equipped with a single antenna. The STAR-RIS has $N$ elements, each of which can independently change the amplitudes and phase shifts of reflected/transmitted signals upon it, thus reconfiguring the signal propagation proactively. Let $\sqrt{\beta^{\text{t}}_{n}}e^{j\theta^{\text{t}}_{n}}$ and $\sqrt{\beta^{\text{r}}_{n}}e^{j\theta^{\text{r}}_{n}}$ denote the transmission and reflection coefficients of $n$th element of STAR-RIS, where $\beta^{\text{t}}_{n}+\beta^{\text{r}}_{n}\leq 1$\footnote{The transmission and reflection amplitude coefficients should obey the law of energy conservation \cite{Liu_magazine}.}, and $\theta^{\text{t}}_{n}, \theta^{\text{r}}_{n}\in(0,2\pi]$ \cite{X.Mu_STAR-RIS,J.Xu_STAR-RIS}. Additionally, there is a smart controller, connected to the STAR-RIS, communicates with the BS via a separate wireless link for pilot acquisition and instant control instruction interaction \cite{R.Zhang_IRS_magazine,R.Zhang_IRS_passive}.

The quasi-static block fading channels are considered in this paper, which means that all the channels remain constant in one fading block but vary over the different fading blocks. The baseband equivalent channels from STAR-RIS to BS, IU, OU, and E are denoted as $\mathbf{G}\in\mathbb{C}^{N\times M}$, $\mathbf{h}_{\text{I},\text{S}}\in\mathbb{C}^{N\times 1}$, $\mathbf{h}_{\text{O},\text{S}}\in\mathbb{C}^{N\times 1}$ and $\mathbf{h}_{\text{E},\text{S}}\in\mathbb{C}^{N\times 1}$. Since the positions of STAR-RIS and BS can be properly selected to favor line-of-sight (LoS) transmission, the Rician fading model is assumed for $\mathbf{G}$, i.e., $\mathbf{G}=\sqrt{\frac{\kappa}{1+\kappa}}\mathbf{\hat{G}}+\sqrt{\frac{1}
{1+\kappa}}\mathbf{\tilde{G}}$, where $\mathbf{\hat{G}}$ denotes the LoS component, $\mathbf{\tilde{G}}$ denotes the non-line-of-sight (NLoS) component, and $\kappa$ denotes the corresponding Rician factor. While for the remaining links, we consider the Rayleigh fading model for them as the mobile users are usually situated in rich-scattering environments.


Moreover, we assume that the instantaneous CSI of legitimate channels is perfectly available, where the reflecting and transmitting channels can be separately estimated by the anchor-assisted channel estimation approach in \cite{Guan_channel_estimation}. While for E, two types of eavesdropping CSI assumptions are considered in this paper. To elaborate, 1) If E is an internal wiretap node (i.e., another active user of the network), the full eavesdropping CSI is assumed to be available at the legitimate nodes; 2) while if E is an external wiretap node, such as the purely passive eavesdropping device of the third-party network without any pilot information interaction with legitimate nodes, it is assumed that only the statistical eavesdropping CSI is available for legitimate nodes, which can be estimated via the long-term monitoring \cite{L.Yang_adaptive_eve}.


Following the uplink NOMA principle, IU and OU transmit $s_{\text{I}}$ and $s_{\text{O}}$ to the BS via the STAR-RIS, where $s_{\text{I}}$ and  $s_{\text{O}}$ are corresponding signals with $\mathbb{E}\{|s_{\text{I}}|^{2}\}=\mathbb{E}\{|s_{\text{O}}|^{2}\}=1$. Accordingly, the received signals at the BS is given by
\begin{equation}
\label{1}
\mathbf{x}_{\text{B}}=(\mathbf{G}^{H}\bm{\Theta}^{\text{t}}\mathbf{h}_{\text{I},\text{S}})\sqrt{P_{\text{I}}}s_{\text{I}}+
(\mathbf{G}^{H}\bm{\Theta}^{\text{r}}\mathbf{h}_{\text{O},\text{S}})\sqrt{P_{\text{O}}}s_{\text{O}}+\mathbf{n}_{\text{B}},
\end{equation}
where $P_{\text{I}}$ and $P_{\text{O}}$ denote the transmit power at the IU and OU, $\bm{\Theta}^{\text{t}}=\text{diag}([\sqrt{\beta^{\text{t}}_{1}}e^{j\theta^{\text{t}}_{1}},\dots,\sqrt{\beta^{\text{t}}_{N}}e^{j\theta^{\text{t}}_{N}}])$,  $\bm{\Theta}^{\text{r}}=\text{diag}([\sqrt{\beta^{\text{r}}_{1}}e^{j\theta^{\text{r}}_{1}},\dots,\sqrt{\beta^{\text{r}}_{N}}e^{j\theta^{\text{r}}_{N}}])$ denote the transmission and reflection coefficient matrices respectively, and $\mathbf{n}_{\text{B}}\sim \mathcal{CN}(0,\sigma^{2}\mathbf{I})$ represents the additive white Gaussian noise (AWGN) at the BS. On receiving superimposed signals, the multi-antenna BS exploits a receive beamforming vector $\mathbf{w}\in\mathbb{C}^{M\times 1}$ to combine the $\mathbf{x}_{\text{B}}$ linearly by adjusting the phases/amplitudes of the signals \cite{R.Long_receive_beam,R.Long_IRS_SIMO}. Thus, the extracted signal through the receive beamforming $\mathbf{w}$ is expressed as
\begin{align}
\nonumber
y_{\text{B}}&=\mathbf{w}^{H}\mathbf{x}_{\text{B}}\\ \label{2}
&=\mathbf{w}^{H}(\mathbf{G}^{H}\bm{\Theta}^{\text{t}}\mathbf{h}_{\text{I},\text{S}})\sqrt{P_{\text{I}}}s_{\text{I}}+
\mathbf{w}^{H}(\mathbf{G}^{H}\bm{\Theta}^{\text{r}}\mathbf{h}_{\text{O},\text{S}})\sqrt{P_{\text{O}}}s_{\text{O}}+\mathbf{w}^{H}\mathbf{n}_{\text{B}}.
\end{align}
As for E, as it is located near the BS and can intercept the superimposed signals, the received signals are given by
\begin{equation}
\label{3}
y_{\text{E}}=(\mathbf{h}_{\text{E},\text{S}}^{H}\bm{\Theta}^{\text{t}}\mathbf{h}_{\text{I},\text{S}})\sqrt{P_{\text{I}}}s_{\text{I}}+
(\mathbf{h}_{\text{E},\text{S}}^{H}\bm{\Theta}^{\text{r}}\mathbf{h}_{\text{O},\text{S}})\sqrt{P_{\text{O}}}s_{\text{O}}+n_{\text{E}},
\end{equation}
where $n_{\text{E}}\sim \mathcal{CN}(0,\sigma^{2})$ represents the AWGN at the E.

In uplink NOMA transmission, we assume that the BS adopts the SIC technique to decode the superimposed signals with the optimal decoding order, which starts from the stronger signals to the weaker signals \cite{L.Lv_magazine}. To facilitate the SIC decoding order optimization, we use $u_{\text{I}}$ and $u_{\text{O}}$ to represent the decoding order of IU and OU respectively, which satisfies $u_{\text{I}},u_{\text{O}}\in\{0,1\}, u_{\text{I}}+u_{\text{O}}=1$. For instance, if the BS first decodes $s_{\text{I}}$ by treating $s_{\text{O}}$ as interference, we have $u_{\text{I}}=1$. The received signal-to-interference-plus-noise ratio (SINR) at the BS to decode $s_{\text{I}}$ is given by
\begin{equation}
\label{4}
\gamma_{\text{I}}=\frac{P_{\text{I}}|\mathbf{w}^{H}(\mathbf{G}^{H}\bm{\Theta}^{\text{t}}\mathbf{h}_{\text{I},\text{S}})|^2}
{u_{\text{I}}P_{\text{O}}|\mathbf{w}^{H}(\mathbf{G}^{H}\bm{\Theta}^{\text{r}}\mathbf{h}_{\text{O},\text{S}})|^2+\sigma^2\|\mathbf{w}\|^2}=
\frac{P_{\text{I}}|\mathbf{w}^{H}(\mathbf{G}^{H}\bm{\Theta}^{\text{t}}\mathbf{h}_{\text{I},\text{S}})|^2}
{P_{\text{O}}|\mathbf{w}^{H}(\mathbf{G}^{H}\bm{\Theta}^{\text{r}}\mathbf{h}_{\text{O},\text{S}})|^2+\sigma^2\|\mathbf{w}\|^2}.
\end{equation}
Then, the BS can eliminate $s_{\text{I}}$ from its observations and decode $s_{\text{O}}$ without interference. As such, we have $u_{\text{O}}=0$, and the received signal-to-noise ratio (SNR) to decode $s_{\text{O}}$ is given by
\begin{equation}
\label{5}
\gamma_{\text{O}}=\frac{P_{\text{O}}|\mathbf{w}^{H}(\mathbf{G}^{H}\bm{\Theta}^{\text{r}}\mathbf{h}_{\text{O},\text{S}})|^2}
{u_{\text{O}}P_{\text{I}}|\mathbf{w}^{H}(\mathbf{G}^{H}\bm{\Theta}^{\text{t}}\mathbf{h}_{\text{I},\text{S}})|^2+\sigma^2\|\mathbf{w}\|^2}=
\frac{P_{\text{O}}|\mathbf{w}^{H}(\mathbf{G}^{H}\bm{\Theta}^{\text{r}}\mathbf{h}_{\text{O},\text{S}})|^2}
{\sigma^2\|\mathbf{w}\|^2}.
\end{equation}
While for the other case that the BS first decodes $s_{\text{O}}$, we have $u_{\text{I}}=0$ and $u_{\text{O}}=1$.

As for E, it has different channel conditions from the BS, which may lead to the different SIC decoding orders at E from BS. In other words, both $s_{\text{I}}$ and $s_{\text{O}}$ might be decoded at the second SIC stage, which enjoys an interference-free decoding. To avoid the secrecy performance loss caused by the uncertain decoding order at E, we consider the worst-case assumption from the perspective of PLS in this paper, i.e., E decodes both two signals without suffering from co-channel interference. Thus, the received SNRs at E to decode $s_{\text{I}}$ and $s_{\text{O}}$ are given by
\begin{equation}
\label{6}
\gamma_{\text{E},\text{I}}=\frac{P_{\text{I}}|\mathbf{h}_{\text{E},\text{S}}^{H}\bm{\Theta}^{\text{t}}\mathbf{h}_{\text{I},\text{S}}|^2}
{\sigma^2},\quad \gamma_{\text{E},\text{O}}=\frac{P_{\text{O}}|\mathbf{h}_{\text{E},\text{S}}^{H}\bm{\Theta}^{\text{r}}\mathbf{h}_{\text{O},\text{S}}|^2}
{\sigma^2}.
\end{equation}

\vspace{-5mm}
\subsection{Problem Formulation for Full Eavesdropping CSI}
\vspace{-1mm}
In the full eavesdropping CSI case, as the legitimate nodes possess the instantaneous channel gains of all the transmission links, we can evaluate the channel capacities of the legitimate and wiretap channels. Therefore, we consider the adaptive-rate wiretap code scheme \cite{Wyner}, that elaborately design the codeword rate $R_{\text{c},\rho}=C_{\rho}=\log_{2}(1+\gamma_{\rho})$ and redundant rate $R_{\text{r},\rho}=C_{\text{E},\rho}=\log_{2}(1+\gamma_{\text{E},\rho})$ ($\rho\in\{\text{I},\text{O}\}$) to minimize the mutual information between the IU/OU and E, thus securing legitimate transmissions perfectly. Accordingly, the secrecy capacity is given by
\begin{equation}
\label{7}
R_{\text{s},\rho} = [C_{\rho}-C_{\text{E},\rho}]^{+}, \rho\in\{\text{I},\text{O}\},
\end{equation}
where $[\cdot]^{+}=\text{max}(\cdot,0)$. In this case, we aim to maximize the minimum secrecy capacity among IU and OU by jointly optimizing transmit power at the IU/OU, receive beamforming at the BS, and passive beamforming at the STAR-RIS, subject to the total power consumption at the IU/OU, SIC decoding order constraint at the BS, and the phase shifts/amplitudes constraints at the STAR-RIS. The optimization problem is formulated as follows.\vspace{-2mm}
\begin{subequations}
\begin{align}
\label{8a} &\max\limits_{P_{\text{I}},P_{\text{O}},\bm{\Theta}^{\text{t}},\bm{\Theta}^{\text{r}},\mathbf{w}}\quad \min\limits_{\rho\in\{\text{I},\text{O}\}}\  R_{\text{s},\rho},\\
\label{8b}&\quad\text{s.t.} \quad P_{\text{I}} \leq P_{\text{I},\text{max}},\quad P_{\text{O}} \leq P_{\text{O},\text{max}},\\
\label{8c}&\quad\quad\,\  \begin{cases}
P_{\text{O}}|\mathbf{w}^{H}(\mathbf{G}^{H}\bm{\Theta}^{\text{r}}\mathbf{h}_{\text{O},\text{S}})|^2\leq P_{\text{I}}|\mathbf{w}^{H}(\mathbf{G}^{H}\bm{\Theta}^{\text{t}}\mathbf{h}_{\text{I},\text{S}})|^2, \quad \text{if $u_{\text{I}}=1$, $u_{\text{O}}=0$,}\\
P_{\text{I}}|\mathbf{w}^{H}(\mathbf{G}^{H}\bm{\Theta}^{\text{t}}\mathbf{h}_{\text{I},\text{S}})|^2\leq P_{\text{O}}|\mathbf{w}^{H}(\mathbf{G}^{H}\bm{\Theta}^{\text{r}}\mathbf{h}_{\text{O},\text{S}})|^2, \quad \text{if $u_{\text{I}}=0$, $u_{\text{O}}=1$,}
\end{cases} \\
\label{8d}&\quad\quad\quad\,\, \|\mathbf{w}\|^{2}= 1,\\
\label{8e}&\quad\quad\quad\,\, \theta^{\text{t}}_{n}, \theta^{\text{r}}_{n}\in[0,2\pi], \ 1\leq n\leq N,\\
\label{8f}&\quad\quad\quad\,\, 0 \leq \beta^{\text{t}}_{n}\leq 1, \quad 0 \leq \beta^{\text{r}}_{n}\leq 1,\quad \beta^{\text{r}}_{n}+\beta^{\text{t}}_{n}\leq 1,\quad 1\leq n\leq N,
\end{align}
\end{subequations}
where $P_{\text{I},\text{max}}$ and $P_{\text{O},\text{max}}$ represent the maximal transmit power at the IU and OU. In problem (8), \eqref{8b} denotes the total power consumption limit at IU/OU; \eqref{8c} denotes the SIC decoding order constraint at the BS; \eqref{8d} denotes the normalization constraint of the receive beamforming $\mathbf{w}$; \eqref{8e} and \eqref{8f} denote the amplitudes and phase shifts constraints at the STAR-RIS, respectively. Furthermore, to avoid the secrecy performance loss caused by the SIC decoding order, we solve the problem (8) for the cases of $u_{\text{I}}=0,u_{\text{O}}=1$ and $u_{\text{I}}=1,u_{\text{O}}=0$ and select the maximal secrecy capacity as the final solution.
\vspace{-4mm}
\subsection{Problem Formulation for Statistical Eavesdropping CSI}\vspace{-1mm}
In statistical eavesdropping CSI case, since there is no available instantaneous channel information of the wiretap links, it is impossible to exactly evaluate the channel capacity of E, which invalidates the adaptive-rate wiretap code scheme above. To guarantee the perfect secure transmission of the legitimate users, the constant-rate wiretap code scheme is considered under this circumstance, where IU and OU transmit signals with the constant codeword rate $R_{\text{c},\rho}\leq C_{\rho}=\log_{2}(1+\gamma_{\rho})$ and the constant secrecy rate $R_{\text{s},\rho}$ ($\rho\in\{\text{I},\text{O}\}$) \cite{M.Bloch_TIT}. Ulteriorly, we introduce the SOP as the security metric for network secrecy performance assessment, which is defined as the probability that the positive difference between the presupposed codeword rate $R_{\text{c},\rho}$  and eavesdropping channel capacity $C_{\text{E},\rho}=\log_{2}(1+\gamma_{\text{E},\rho})$ is less than the target secrecy rate $R_{\text{s},\rho}$. It is given by
\begin{equation}
\label{9}
P_{\text{SOP},\rho}=\mathbb{P}\left(R_{\text{c},\rho}-C_{\text{E},\rho}<R_{\text{s},\rho}\right).
\end{equation}

We aim to minimize the maximum SOP of legitimate users subject to the total power consumption and codeword rate constraint at the IU/OU, SIC decoding order constraint at the BS, and the phase shift/amplitude constraints at the STAR-RIS, by jointly optimizing transmit power at the IU/OU, receive beamforming at the BS and passive beamforming at the STAR-RIS. The problem is formulated as follows.\vspace{-2mm}
\begin{subequations}
\begin{align}
\label{10a} &\min\limits_{P_{\text{I}},P_{\text{O}},\bm{\Theta}^{\text{t}},\bm{\Theta}^{\text{r}},\mathbf{w}}\quad \max\limits_{\rho\in\{\text{I},\text{O}\}}\ P_{\text{SOP},\rho},\\
\label{10b}&\quad\text{s.t.} \quad\ \  \eqref{8b}-\eqref{8f},\\
\label{10c}&\quad\quad\quad\,\, R_{\text{c},\rho}\leq \log_{2}(1+\gamma_{\rho}),\quad \rho\in\{\text{I},\text{O}\},
\end{align}
\end{subequations}
where \eqref{10c} denotes the constant codeword rate. 

\begin{remark}
    Due to the two-layer coupling, the formulated problems (8) and (10) are more challenging to tackle than the conventional beamforming optimization problem of the reflecting-only RIS assisted downlink transmission. Specifically, 1) in the outer layer, A three-variable coupling exists in the objective function (8a) and constraints (8c), which cannot be coped with traditional methods applied in joint active and passive beamforming design \cite{R.Zhang_IRS_passive,M.Cui_IRS_PLS,B.Ning_MIMO}; 2) in the inner layer, the amplitudes of transmission and reflection beamforming are restricted by each other because of the law of energy conservation, which further exacerbates the optimization difficulty. In the next two sections, we first propose an AHB algorithm to handle the problem (8) and then extend it to the problem (10).
\end{remark}

\vspace{-2mm}
\section{Proposed Solution for Full Eavesdropping CSI}

In this section, we develop an efficient AHB algorithm to maximize the minimum secrecy capacity of IU and OU, where a two-layer iterative algorithm is proposed to optimize reflection/transmission coefficients and receive beamforming jointly, while the optimal transmit power policy is derived in the closed-form expression.

\vspace{-5mm}
\subsection{Joint Beamforming Optimization}

Define $\mathbf{u}_{\text{t}}=[\sqrt{\beta^{\text{t}}_{1}}e^{j\theta^{\text{t}}_{1}},\dots,\sqrt{\beta^{\text{t}}_{N}}e^{j\theta^{\text{t}}_{N}}]^{T}$, $\mathbf{u}_{\text{r}}=[\sqrt{\beta^{\text{r}}_{1}}e^{j\theta^{\text{r}}_{1}},\dots,\sqrt{\beta^{\text{r}}_{N}}e^{j\theta^{\text{r}}_{N}}]^{T}$,
$\mathbf{q}_{\rho}=\mathbf{G}^{H}\text{diag}(\mathbf{h}_{\rho,\text{S}})$ and $\mathbf{q}_{\text{E},\rho}=\mathbf{h}_{\text{E},\text{S}}\odot\mathbf{h}_{\rho,\text{S}}$, $\rho\in\{\text{I},\text{O}\}$. With the fixed $P_{\text{I}}$ and $P_{\text{O}}$, we rewrite the problem (8) as
\begin{subequations}
\begin{align}
\label{11a} &\max\limits_{\mathbf{U}_{\text{t}},\mathbf{U}_{\text{r}},\mathbf{W}}\quad \min\limits_{\rho\in\{\text{I},\text{O}\}}\  R_{\text{s},\rho},\\
\label{11b}&\quad\text{s.t.}   \begin{cases}
P_{\text{O}}\text{Tr}(\mathbf{q}_{\text{O}}^{H}\mathbf{W}\mathbf{q}_{\text{O}}\mathbf{U}_{\text{r}})\leq P_{\text{I}}\text{Tr}(\mathbf{q}_{\text{I}}^{H}\mathbf{W}\mathbf{q}_{\text{I}}\mathbf{U}_{\text{t}}), \quad \text{if $u_{\text{I}}=1$, $u_{\text{O}}=0$,}\\
P_{\text{I}}\text{Tr}(\mathbf{q}_{\text{I}}^{H}\mathbf{W}\mathbf{q}_{\text{I}}\mathbf{U}_{\text{t}})\leq P_{\text{O}}\text{Tr}(\mathbf{q}_{\text{O}}^{H}\mathbf{W}\mathbf{q}_{\text{O}}\mathbf{U}_{\text{r}}), \quad \text{if $u_{\text{I}}=0$, $u_{\text{O}}=1$,}
\end{cases} \\
\label{11c}&\quad\quad\quad\,\, {\text{Tr}}(\mathbf{W})= 1,\\
\label{11d}&\quad\quad\quad\,\, \mathbf{U}_{\text{t}}(n,n)+\mathbf{U}_{\text{r}}(n,n) \leq 1, \quad \ 1\leq n\leq N,\\
\label{11e}&\quad\quad\quad\,\, \mathbf{W}\succeq \mathbf{0}, \quad \mathbf{U}_{\text{t}}\succeq\mathbf{0}, \quad \mathbf{U}_{\text{r}}\succeq\mathbf{0},\\
\label{11f}&\quad\quad\quad\,\,\mathbf{U}_{\text{t}}=\mathbf{u}_{\text{t}}\mathbf{u}_{\text{t}}^{H}, \quad \mathbf{U}_{\text{r}}=\mathbf{u}_{\text{r}}\mathbf{u}_{\text{r}}^{H}, \quad \mathbf{W}=\mathbf{w}\mathbf{w}^{H},
\end{align}
\end{subequations}
where $\gamma_{\text{E},\text{I}}=\frac{P_{\text{I}}\text{Tr}(\mathbf{q}_{\text{E},\text{I}}\mathbf{q}_{\text{E},\text{I}}^{H}\mathbf{U}_{\text{t}})}
{\sigma^2}$ and $\gamma_{\text{E},\text{O}}=\frac{P_{\text{O}}\text{Tr}(\mathbf{q}_{\text{E},\text{O}}\mathbf{q}_{\text{E},\text{O}}^{H}\mathbf{U}_{\text{r}})}
{\sigma^2}$. To tackle the problem (11) with non-convex terms \eqref{11a}, \eqref{11b} and \eqref{11f}, we adopt the polarization identity \cite{X.-D_Polarization} to decouple the reflection/transmission coefficients of STAR-RIS and the receive beamforming at the BS. Specifically, according to \cite[eq.(1.4.50)]{X.-D_Polarization}, we can equivalently rewrite $\text{Tr}(\mathbf{q}_{\rho}^{H}\mathbf{W}\mathbf{q}_{\rho}\mathbf{U}_{\text{t}/\text{r}})$ as
    \begin{equation}\label{12}
        \text{Tr}(\mathbf{q}_{\rho}^{H}\mathbf{W}\mathbf{q}_{\rho}\mathbf{U}_{\text{t}/\text{r}})=\frac{1}{4}\Big(
        \underbrace{\|\mathbf{q}_{\rho}^{H}\mathbf{W}\mathbf{q}_{\rho}+\mathbf{U}_{\text{t}/\text{r}}\|_{F}^{2}}_{a_{1}}-
        \underbrace{\|\mathbf{q}_{\rho}^{H}\mathbf{W}\mathbf{q}_{\rho}-\mathbf{U}_{\text{t}/\text{r}}\|_{F}^{2}}_{a_{2}}\Big).
    \end{equation}
    where $\mathbf{U}_{\text{t}/\text{r}}=\mathbf{U}_{\text{t}}$ in the case of $\rho=\text{I}$ while $\mathbf{U}_{\text{t}/\text{r}}=\mathbf{U}_{\text{r}}$ in the case of $\rho=\text{O}$. By leveraging the first-order Taylor expansion of $a_{1}$ and $a_{2}$, the convex upper bound and the concave lower bound expressions of $\text{Tr}(\mathbf{q}_{\rho}\mathbf{W}\mathbf{q}_{\rho}^{H}\mathbf{U}_{\text{t}/\text{r}})$ are respectively given by
    \begin{align}\nonumber
         2\text{Tr}\Big(
        &(\mathbf{q}_{\rho}^{H}\mathbf{W}\mathbf{q}_{\rho}+\mathbf{U}_{\text{t}/\text{r}})
        (\mathbf{q}_{\rho}^{H}\mathbf{\widetilde{W}}\mathbf{q}_{\rho}+\mathbf{\widetilde{U}}_{\text{t}/\text{r}})^{H}\Big)-\\ \label{13}
        &\|\mathbf{q}_{\rho}^{H}\mathbf{\widetilde{W}}\mathbf{q}_{\rho}+\mathbf{\widetilde{U}}_{\text{t}/\text{r}}\|_{F}^{2}- \|\mathbf{q}_{\rho}^{H}\mathbf{W}\mathbf{q}_{\rho}-\mathbf{U}_{\text{t}/\text{r}}\|_{F}^{2}\geq4\mathfrak{T}_{\rho,\text{t}/\text{r}}^{\text{lower}},
    \end{align}
    \begin{align}\nonumber
         4\mathfrak{T}_{\rho,\text{t}/\text{r}}^{\text{upper}}\geq &-2\text{Tr}\Big(
        (\mathbf{q}_{\rho}^{H}\mathbf{W}\mathbf{q}_{\rho}-\mathbf{U}_{\text{t}/\text{r}})
        (\mathbf{q}_{\rho}^{H}\mathbf{\widetilde{W}}\mathbf{q}_{\rho}-\mathbf{\widetilde{U}}_{\text{t}/\text{r}})^{H}\Big)+\\ \label{14}
        &\|\mathbf{q}_{\rho}^{H}\mathbf{\widetilde{W}}\mathbf{q}_{\rho}+\mathbf{\widetilde{U}}_{\text{t}/\text{r}}\|_{F}^{2}+
         \|\mathbf{q}_{\rho}^{H}\mathbf{W}\mathbf{q}_{\rho}+\mathbf{U}_{\text{t}/\text{r}}\|_{F}^{2}.
    \end{align}
    where $\mathbf{\widetilde{W}}$ and $\mathbf{\widetilde{U}}_{\text{t}/\text{r}}$ denote the local points generated in the previous iteration, while $\mathfrak{T}_{\rho,\text{t}/\text{r}}^{\text{lower}}$ and $\mathfrak{T}_{\rho,\text{t}/\text{r}}^{\text{upper}}$ are slack variables introduced to represent the lower bound and upper bound of $\text{Tr}(\mathbf{q}_{\rho}^{H}\mathbf{W}\mathbf{q}_{\rho}\mathbf{U}_{\text{t}/\text{r}})$, respectively.

With the results above, the SIC decoding order constraints \eqref{11b} can be conservatively transformed into the following linear form
\begin{equation}\label{15}
    \begin{cases}
P_{\text{O}}\mathfrak{T}_{\text{O},\text{r}}^{\text{upper}}\leq P_{\text{I}}\mathfrak{T}_{\text{I},\text{t}}^{\text{lower}}, \quad \text{if $u_{\text{I}}=1$, $u_{\text{O}}=0$,}\\
P_{\text{I}}\mathfrak{T}_{\text{I},\text{t}}^{\text{upper}}\leq P_{\text{O}}\mathfrak{T}_{\text{O},\text{r}}^{\text{lower}}, \quad \text{if $u_{\text{I}}=0$, $u_{\text{O}}=1$.}
\end{cases}
\end{equation}
Similarly, the lower bound of the secrecy capacity of IU/OU can be expressed as $R_{\text{s},\rho}^{\text{lower}}=\log_{2}(1+\gamma_{\rho}^{\text{lower}})-\log_{2}(1+\gamma_{\text{E},\rho})$, where
\begin{equation}\label{16}
\begin{cases}
 \gamma_{\text{I}}\geq \gamma_{\text{I}}^{\text{lower}}=\frac{P_{\text{I}}\mathfrak{T}_{\text{I},\text{t}}^{\text{lower}}}
{P_{\text{O}}\mathfrak{T}_{\text{O},\text{r}}^{\text{upper}}+\sigma^2},\ \  \gamma_{\text{O}}\geq \gamma_{\text{O}}^{\text{lower}}=\frac{P_{\text{O}}\mathfrak{T}_{\text{O},\text{r}}^{\text{lower}}}
{\sigma^2}, \quad \text{if $u_{\text{I}}=1$, $u_{\text{O}}=0$,}\\
\gamma_{\text{I}}\geq \gamma_{\text{I}}^{\text{lower}}=\frac{P_{\text{I}}\mathfrak{T}_{\text{I},\text{t}}^{\text{lower}}}
{\sigma^2},\ \  \gamma_{\text{O}}\geq \gamma_{\text{O}}^{\text{lower}}=\frac{P_{\text{O}}\mathfrak{T}_{\text{O},\text{r}}^{\text{lower}}}
{P_{\text{I}}\mathfrak{T}_{\text{I},\text{t}}^{\text{upper}}+\sigma^2}, \quad \text{if $u_{\text{I}}=0$, $u_{\text{O}}=1$.}
\end{cases}
\end{equation}
Therefore, the optimization problem (11) can be reformulated as
\begin{subequations}
\begin{align}
\label{17a} &\max\limits_{\mathbf{U}_{\text{t}},\mathbf{U}_{\text{r}},\mathbf{W},\mathfrak{T}_{\rho,\text{t}/\text{r}}^{\text{lower}},
\mathfrak{T}_{\rho,\text{t}/\text{r}}^{\text{upper}},R_{\text{s},\text{min}}}\quad \min\limits_{\rho\in\{\text{I},\text{O}\}}\  \frac{1+\gamma_{\rho}^{\text{lower}}}{1+\gamma_{\text{E},\rho}},\\
\label{17b}&\quad\text{s.t.} \quad\ \eqref{11c}-\eqref{11e},\eqref{13}-\eqref{16},\\
\label{17c}&\quad\quad\quad\,\,\,\text{rank}(\mathbf{U}_{\text{t}})=1, \quad \text{rank}(\mathbf{U}_{\text{r}})=1, \quad \text{rank}(\mathbf{W})=1.
\end{align}
\end{subequations}
The problem (17) remains non-convex and challenging to deal with due to the non-convexity in \eqref{16}, \eqref{17a} and \eqref{17c}. To overcome this issue, we introduce slack variable $\varphi$ to denote the lower bound SINR of the strong signal, i.e.,
\begin{equation}\label{18}
\begin{cases}
 P_{\text{I}}\mathfrak{T}_{\text{I},\text{t}}^{\text{lower}}\geq
(P_{\text{O}}\mathfrak{T}_{\text{O},\text{r}}^{\text{upper}}+\sigma^2)\varphi, \quad \text{if $u_{\text{I}}=1$, $u_{\text{O}}=0$,}\\
 P_{\text{O}}\mathfrak{T}_{\text{O},\text{r}}^{\text{lower}}\geq
(P_{\text{I}}\mathfrak{T}_{\text{I},\text{t}}^{\text{upper}}+\sigma^2)\varphi, \quad \text{if $u_{\text{I}}=0$, $u_{\text{O}}=1$.}
\end{cases}
\end{equation}
Then, we resort to Lemma 1 to construct the convex upper approximation function of right-hand side of \eqref{18} as follows
\begin{equation}\label{19}
 \widehat{H}(\mathfrak{T}_{\rho,\text{t}/\text{r}}^{\text{upper}},\varphi,\varpi_{\rho,\text{t}/\text{r}})=\frac{1}{2}\left(
 \varpi_{\rho,\text{t}/\text{r}}(P_{\rho}\mathfrak{T}_{\rho,\text{t}/\text{r}}^{\text{upper}}+\sigma^2)^{2}+\frac{\varphi^{2}}{\varpi_{\rho,\text{t}/\text{r}}}\right),
\end{equation}
where $\varpi_{\rho,\text{t}/\text{r}}=\frac{\varphi}{P_{\rho}\mathfrak{T}_{\rho,\text{t}/\text{r}}^{\text{upper}}+\sigma^2}$.

\begin{lemma}
    According to \cite{A.Beck_SPCA}, the convex upper approximation function  $\widehat{H}(\mathfrak{T}_{\rho,\text{t}/\text{r}}^{\text{upper}},\varphi,\varpi_{\rho,\text{t}/\text{r}})$ of the function $H(\mathfrak{T}_{\rho,\text{t}/\text{r}}^{\text{upper}},\varphi)=(P_{\rho}\mathfrak{T}_{\rho,\text{t}/\text{r}}^{\text{upper}}+\sigma^2)\varphi$ for $\rho\in\{\text{I},\text{O}\}$ is supposed to meet: 1) for arbitrary given parameter $\varpi_{\rho,\text{t}/\text{r}}\neq0$, $\widehat{H}(\mathfrak{T}_{\rho,\text{t}/\text{r}}^{\text{upper}},\varphi,\varpi_{\rho,\text{t}/\text{r}})\geq
    H(\mathfrak{T}_{\rho,\text{t}/\text{r}}^{\text{upper}},\varphi)$ holds; 2) for arbitrary positive $\mathfrak{T}_{\rho,\text{t}/\text{r}}^{\text{upper}}$ and $\varphi$, there always exists a parameter $\varpi_{\rho,\text{t}/\text{r}}$, which satisfies
    \begin{subequations}
    \begin{align}
    \label{20a} \widehat{H}(\mathfrak{T}_{\rho,\text{t}/\text{r}}^{\text{upper}},\varphi,\varpi_{\rho,\text{t}/\text{r}})&=
    H(\mathfrak{T}_{\rho,\text{t}/\text{r}}^{\text{upper}},\varphi),\\
    \label{20b} \nabla_{\mathfrak{T}_{\rho,\text{t}/\text{r}}^{\text{upper}}}\widehat{H}(\mathfrak{T}_{\rho,\text{t}/\text{r}}^{\text{upper}},\varphi,\varpi_{\rho,\text{t}/\text{r}})&=
    \nabla_{\mathfrak{T}_{\rho,\text{t}/\text{r}}^{\text{upper}}} H(\mathfrak{T}_{\rho,\text{t}/\text{r}}^{\text{upper}},\varphi),\\
    \label{20c} \nabla_{\varphi}\widehat{H}(\mathfrak{T}_{\rho,\text{t}/\text{r}}^{\text{upper}},\varphi,\varpi_{\rho,\text{t}/\text{r}})&=
    \nabla_{\varphi} H(\mathfrak{T}_{\rho,\text{t}/\text{r}}^{\text{upper}},\varphi).
    \end{align}
    \end{subequations}
\end{lemma}

As a result, we are capable of rewriting \eqref{17a} as
\begin{equation}\label{21}
\min\ \left\{ \frac{1+\varphi}{1+u_{\text{I}}\gamma_{\text{E},\text{I}}+u_{\text{O}}\gamma_{\text{E},\text{O}}},
 \frac{1+u_{\text{I}}\frac{P_{\text{O}}\mathfrak{T}_{\text{O},\text{r}}^{\text{lower}}}
{\sigma^2}+u_{\text{O}}\frac{P_{\text{I}}\mathfrak{T}_{\text{I},\text{r}}^{\text{lower}}}
{\sigma^2}}{1+u_{\text{I}}\gamma_{\text{E},\text{O}}
 +u_{\text{O}}\gamma_{\text{E},\text{I}}}\right\},
\end{equation}
where $\varphi$ satisfies
\begin{equation}\label{22}
 \begin{cases}
 P_{\text{I}}\mathfrak{T}_{\text{I},\text{t}}^{\text{lower}}\geq
\widehat{H}(\mathfrak{T}_{\text{O},\text{r}}^{\text{upper}},\varphi,\varpi_{\text{O},\text{r}}), \quad \text{if $u_{\text{I}}=1$, $u_{\text{O}}=0$,}\\
 P_{\text{O}}\mathfrak{T}_{\text{O},\text{r}}^{\text{lower}}\geq
\widehat{H}(\mathfrak{T}_{\text{I},\text{t}}^{\text{upper}},\varphi,\varpi_{\text{I},\text{t}}), \quad \text{if $u_{\text{I}}=0$, $u_{\text{O}}=1$.}
\end{cases}
\end{equation}
Moreover, to guarantee rank-one property of $\mathbf{U}_{\text{t}/\text{r}}$, we adopt the difference-of-convex (DC) relaxation method \cite{T.Jiang_DCP} to extract the rank-one solution from high-rank matrix. To elaborate, $\text{rank}(\mathbf{U}_{\text{t}/\text{r}})=1$ is equivalent to
$\text{Tr}(\mathbf{U}_{\text{t}/\text{r}})=\|\mathbf{U}_{\text{t}/\text{r}}\|_{2}$, where $\text{Tr}(\mathbf{U}_{\text{t}/\text{r}})=\sum\nolimits_{i=1}^{N}\sigma_{i}$ and $\|\mathbf{U}_{\text{t}/\text{r}}\|_{2}=\sigma_{1}$. With
\cite[Prop. 2]{T.Jiang_DCP}, it can be further relaxed into the DC constraint below
\begin{equation}
\label{23}
\Re(\text{Tr}(\mathbf{U}_{\text{t}/\text{r}}^{H}(\mathbf{I}-\mathbf{u}_{\text{t}/\text{r},1}\mathbf{u}_{\text{t}/\text{r},1}^{H})))\leq\varrho_{\text{t}/\text{r}},
\end{equation}
where $\mathbf{u}_{\text{t}/\text{r},1}\in\mathbb{C}^{N\times 1}$ denotes the leading eigenvector of $\mathbf{U}_{\text{t}/\text{r}}$ obtained in the previous iteration, and $\varrho_{\text{t}/\text{r}}\rightarrow0$ is the penalty factor. Up to this point, the problem (19) can be reformulated as a standard fractional programming with non-convex constraint $\text{rank}(\mathbf{W})=1$, which, however, can be efficiently handled by the Dinkelbach approach \cite{Dinkelbach} and following lemma. The transformed rank-relaxed problem is given by\vspace{-3mm}
\begin{subequations}
\begin{align}
\label{24a} &\max\limits_{\mathbf{U}_{\text{t}},\mathbf{U}_{\text{r}},\mathbf{W},\mathfrak{T}_{\rho,\text{t}/\text{r}}^{\text{lower}},
\atop
\mathfrak{T}_{\rho,\text{t}/\text{r}}^{\text{upper}},\varphi,\varrho_{\text{t}/\text{r}},\xi}\quad  \xi-\tau(\varrho_{\text{t}}+\varrho_{\text{r}}),\\
\label{24b}&\text{s.t.} \ \eqref{11c}-\eqref{11e},\eqref{13}-\eqref{15},\eqref{22},\eqref{23},\\
\label{24c}&\quad\,\,\,1+\varphi-\mu(1+u_{\text{I}}\gamma_{\text{E},\text{I}}+u_{\text{O}}\gamma_{\text{E},\text{O}})\geq\xi,\\
\label{24d}&\quad\,\,\,1+u_{\text{I}}\frac{P_{\text{O}}\mathfrak{T}_{\text{O},\text{r}}^{\text{lower}}}
{\sigma^2}+u_{\text{O}}\frac{P_{\text{I}}\mathfrak{T}_{\text{I},\text{r}}^{\text{lower}}}
{\sigma^2}-\mu(1+u_{\text{I}}\gamma_{\text{E},\text{O}}+u_{\text{O}}\gamma_{\text{E},\text{I}})\geq\xi,
\end{align}
\end{subequations}
where $\tau>0$ is the scaling factor of the penalty terms $\varrho_{\text{t}}$ and $\varrho_{\text{r}}$. Note that $\mu$ denotes the update coefficient, which starts from $0$, and is updated by $\mu\!=\!\min \{ \frac{1+\varphi}{1+u_{\text{I}}\gamma_{\text{E},\text{I}}+u_{\text{O}}\gamma_{\text{E},\text{O}}},\!
 \frac{1+\frac{u_{\text{I}}P_{\text{O}}\mathfrak{T}_{\text{O},\text{r}}^{\text{lower}}}
{\sigma^2}+\frac{u_{\text{O}}P_{\text{I}}\mathfrak{T}_{\text{I},\text{r}}^{\text{lower}}}
{\sigma^2}}{1+u_{\text{I}}\gamma_{\text{E},\text{O}}
 +u_{\text{O}}\gamma_{\text{E},\text{I}}}\}$ in each iteration, while $\xi\geq0$ is an auxiliary variable, which is introduced to measure the approximation gap between $\mu$ and $\min \{ \frac{1+\varphi}{1+u_{\text{I}}\gamma_{\text{E},\text{I}}+u_{\text{O}}\gamma_{\text{E},\text{O}}},\!
 \frac{1+u_{\text{I}}\frac{P_{\text{O}}\mathfrak{T}_{\text{O},\text{r}}^{\text{lower}}}
{\sigma^2}+u_{\text{O}}\frac{P_{\text{I}}\mathfrak{T}_{\text{I},\text{r}}^{\text{lower}}}
{\sigma^2}}{1+u_{\text{I}}\gamma_{\text{E},\text{O}}
 +u_{\text{O}}\gamma_{\text{E},\text{I}}}\}$.
\begin{lemma}
    The obtained solution for the rank-relaxed problem (24) can always guarantee that $\text{rank}(\mathbf{W})=1$.
\end{lemma}
\begin{IEEEproof}
See Appendix A.
\end{IEEEproof}
\begin{remark}
    For any given $\{u_{\text{I}},u_{\text{O}}\}$, the rank-relaxed problem (24) is jointly convex with respect to the optimization variables and can be directly solved by the convex solver (e.g., CVX) in each iteration. Nevertheless, it is supposed to be much careful about the initialization of the value of $\tau$ due to the following reasons. 1) If we initialize scaling factor to be a large value (i.e., $\tau\rightarrow+\infty$), any tiny decrease in the penalty term (i.e., $\varrho_{\text{t}}+\varrho_{\text{r}}$) will lead to a significant increase in the objective function. Thus, the objective function \eqref{24a} will be dominated by the penalty terms, which efficiently guarantees the rank-one property of $\{\mathbf{U}_{\text{t}},\mathbf{U}_{\text{r}}\}$ but neglects our goal of secrecy capacity maximization. 2) In contrast, if we initialize scaling factor to be a small value (i.e., $\tau\rightarrow0$), the secrecy rate penalty caused by the high-rank beamforming matrices becomes trivial, which is conducive to obtain a good starting point for secrecy capacity maximization, but cannot ensure the rank-one property of $\{\mathbf{U}_{\text{t}},\mathbf{U}_{\text{r}}\}$. To overcome this issue, we initialize $\tau$ to be a small value and adopt the two-layer algorithm framework, where $\tau$ remains constant in the inner layer iteration while increasing over the outer layer iterations to converge the stationary point solution of the original problem \cite{Q.T_stationary}. The details of the two-layer iterative algorithm is summarized in \textbf{Algorithm-1}, where $\xi_{\text{th}}$ and $\varrho_{\text{th}}$ denote the convergence accuracy of inner and outer layer loops.
\end{remark}
\vspace{-4mm}
\begin{table}[t]
    \centering
    \begin{tabular}{p{450pt}}
    \toprule
    \textbf{Algorithm-1:} Two-layer Iterative Algorithm \\
    \midrule
    1: \textbf{Initialization}: Initialize the iteration parameters as $\mathbf{\widetilde{W}}(n)$, $\mathbf{\widetilde{U}}_{\text{t}/\text{r}}(n)$, $\varpi_{\rho,\text{t}/\text{r}}(n)$, $\mathbf{u}_{\text{t}/\text{r},1}(n)$, $\mu(n)$ and $\tau(n)$ with $n=1$;\\
    \hangafter 1 
    \hangindent 2em 
    2: \textbf{Outer layer: repeat }\\
    3:          \quad \textbf{Inner layer: repeat }\\
    4:          \qquad For given $\{\mathbf{\widetilde{W}}(n),\mathbf{\widetilde{U}}_{\text{t}/\text{r}}(n),\varpi_{\rho,\text{t}/\text{r}}(n),\mathbf{u}_{\text{t}/\text{r},1}(n),\mu(n),
    \tau(n)\}$, $\{P_{\text{I}}, P_{\text{O}}\}$ and $\{u_{\text{I}},u_{\text{O}}\}$, solve problem (24);\\
    5:          \qquad Set $n = n+1$;\\
    6:          \qquad Update the iteration parameters $\mathbf{\widetilde{W}}(n)=\mathbf{W}(n-1)$, $\mathbf{\widetilde{U}}_{\text{t}/\text{r}}(n)=
    \mathbf{U}_{\text{t}/\text{r}}(n-1)$, $\varpi_{\rho,\text{t}/\text{r}}(n)=\frac{\varphi(n-1)}{P_{\rho}\mathfrak{T}_{\rho,\text{t}/\text{r}}^{\text{upper}}(n-1)+\sigma^2}$,\\
    \qquad \quad
    $\mu(n)=\min \{\frac{1+\varphi(n-1)}{1+u_{\text{I}}\gamma_{\text{E},\text{I}}(n-1)+u_{\text{O}}\gamma_{\text{E},\text{O}}(n-1)},
\frac{1+u_{\text{I}}\gamma_{\text{O}}^{\text{lower}}(n-1)+u_{\text{O}}\gamma_{\text{I}}^{\text{lower}}(n-1)}
{1+u_{\text{I}}\gamma_{\text{E},\text{O}}(n-1)+u_{\text{O}}\gamma_{\text{E},\text{I}}(n-1)}\}$. Then, $\mathbf{u}_{\text{t}/\text{r},1}(n)$ is replaced by the leading \\
    \qquad \quad eigenvector of $\mathbf{U}_{\text{t}/\text{r},1}(n-1)$;\\
    7: \quad  \textbf{Until} $|\xi(n)-\xi(n-1)|\leq\xi_{\text{th}}$;\\
    8: \quad Set $\tau(n) = k\tau(n)$, $k>1$;\\
    9:\textbf{Until} $\varrho_{\text{t}}(n)+\varrho_{\text{r}}(n)\leq\varrho_{\text{th}}$.\\
    \bottomrule
    \end{tabular}
\end{table}
\subsection{Optimal Transmit Power Policy}
With the fixed $\{\mathbf{W},\mathbf{U}_{\text{t}},\mathbf{U}_{\text{r}}\}$, problem (8) is equivalent to
\begin{subequations}
\begin{align}
\label{25a} &\max\limits_{P_{\text{I}},P_{\text{O}}}\quad \min\left\{\frac{1+\frac{P_{\text{I}}\text{Tr}(\mathbf{q}_{\text{I}}^{H}\mathbf{W}\mathbf{q}_{\text{I}}\mathbf{U}_{\text{t}})}
{u_{\text{I}}P_{\text{O}}\text{Tr}(\mathbf{q}_{\text{O}}^{H}\mathbf{W}\mathbf{q}_{\text{O}}\mathbf{U}_{\text{r}})+\sigma^{2}}}{1+
\frac{P_{\text{I}}\text{Tr}(\mathbf{q}_{\text{E},\text{I}}\mathbf{q}_{\text{E},\text{I}}^{H}\mathbf{U}_{\text{t}})}
{\sigma^2}},
\frac{1+\frac{P_{\text{O}}\text{Tr}(\mathbf{q}_{\text{O}}^{H}\mathbf{W}\mathbf{q}_{\text{O}}\mathbf{U}_{\text{r}})}
{u_{\text{O}}P_{\text{I}}\text{Tr}(\mathbf{q}_{\text{I}}^{H}\mathbf{W}\mathbf{q}_{\text{I}}\mathbf{U}_{\text{t}})+\sigma^{2}}}{1+
\frac{P_{\text{O}}\text{Tr}(\mathbf{q}_{\text{E},\text{O}}\mathbf{q}_{\text{E},\text{O}}^{H}\mathbf{U}_{\text{r}})}{\sigma^2}}\right\},\\
\label{25b}&\quad\text{s.t.} \quad P_{\text{I}} \leq P_{\text{I},\text{max}},\quad P_{\text{O}} \leq P_{\text{O},\text{max}},\\
\label{25c}&\quad\quad\,\  \begin{cases}
P_{\text{O}}\text{Tr}(\mathbf{q}_{\text{O}}^{H}\mathbf{W}\mathbf{q}_{\text{O}}\mathbf{U}_{\text{r}})\leq P_{\text{I}}\text{Tr}(\mathbf{q}_{\text{I}}^{H}\mathbf{W}\mathbf{q}_{\text{I}}\mathbf{U}_{\text{t}}), \quad \text{if $u_{\text{I}}=1$, $u_{\text{O}}=0$,}\\
P_{\text{I}}\text{Tr}(\mathbf{q}_{\text{I}}^{H}\mathbf{W}\mathbf{q}_{\text{I}}\mathbf{U}_{\text{t}})\leq P_{\text{O}}\text{Tr}(\mathbf{q}_{\text{O}}^{H}\mathbf{W}\mathbf{q}_{\text{O}}\mathbf{U}_{\text{r}}), \quad \text{if $u_{\text{I}}=0$, $u_{\text{O}}=1$.}
\end{cases}
\end{align}
\end{subequations}
For the case of $u_{\text{I}}=1,u_{\text{O}}=0$, the objective function \eqref{25a} can be rewritten as $\min\{\mathfrak{O}_{\text{I}}(P_{\text{I}},P_{\text{O}}),\\ \mathfrak{O}_{\text{O}}(P_{\text{O}})\}$, where $\mathfrak{O}_{\text{I}}(P_{\text{I}},P_{\text{O}})$ and $\mathfrak{O}_{\text{O}}(P_{\text{O}})$ satisfy
\begin{gather}\label{26}
    \mathfrak{O}_{\text{I}}(P_{\text{I}},P_{\text{O}})=\frac{\left(P_{\text{I}}\text{Tr}(\mathbf{q}_{\text{I}}^{H}\mathbf{W}\mathbf{q}_{\text{I}}\mathbf{U}_{\text{t}})
    +P_{\text{O}}\text{Tr}(\mathbf{q}_{\text{O}}^{H}\mathbf{W}\mathbf{q}_{\text{O}}\mathbf{U}_{\text{r}})+\sigma^{2}\right)\sigma^{2}}
    {\left(P_{\text{I}}\text{Tr}(\mathbf{q}_{\text{E},\text{I}}\mathbf{q}_{\text{E},\text{I}}^{H}\mathbf{U}_{\text{t}})+\sigma^{2}\right)
    \left(P_{\text{O}}\text{Tr}(\mathbf{q}_{\text{O}}^{H}\mathbf{W}\mathbf{q}_{\text{O}}\mathbf{U}_{\text{r}})+\sigma^{2}\right)},\\ \label{27}
    \mathfrak{O}_{\text{O}}(P_{\text{O}})=\frac{P_{\text{O}}\text{Tr}(\mathbf{q}_{\text{O}}^{H}\mathbf{W}\mathbf{q}_{\text{O}}\mathbf{U}_{\text{r}})+\sigma^{2}
}{P_{\text{O}}\text{Tr}(\mathbf{q}_{\text{E},\text{O}}\mathbf{q}_{\text{E},\text{O}}^{H}\mathbf{U}_{\text{r}})+\sigma^2}.
\end{gather}
As a result, the partial derivatives of $\mathfrak{O}_{\text{I}}(P_{\text{I}},P_{\text{O}})$ are derived as
\begin{gather}\label{28}
   \nabla_{P_{\text{I}}}\mathfrak{O}_{\text{I}}\!=\!\frac{\left(
   \text{Tr}(\mathbf{q}_{\text{I}}^{H}\mathbf{W}\mathbf{q}_{\text{I}}\mathbf{U}_{\text{t}})\sigma^{2}\!-\!
   P_{\text{O}}\text{Tr}(\mathbf{q}_{\text{O}}^{H}\mathbf{W}\mathbf{q}_{\text{O}}\mathbf{U}_{\text{r}})
   \text{Tr}(\mathbf{q}_{E,\text{I}}\mathbf{q}_{E,\text{I}}^{H}\mathbf{U}_{\text{t}})\!-\!
   \text{Tr}(\mathbf{q}_{E,\text{I}}\mathbf{q}_{E,\text{I}}^{H}\mathbf{U}_{\text{t}})\sigma^{2}\right)\sigma^{2}}{
   \left(P_{\text{I}}\text{Tr}(\mathbf{q}_{E,\text{I}}\mathbf{q}_{E,\text{I}}^{H}\mathbf{U}_{\text{t}})+\sigma^{2}\right)^2
    \left(P_{\text{O}}\text{Tr}(\mathbf{q}_{\text{O}}^{H}\mathbf{W}\mathbf{q}_{\text{O}}\mathbf{U}_{\text{r}})+\sigma^{2}\right)},\\ \label{29}
    \nabla_{P_{\text{O}}}\mathfrak{O}_{\text{I}}\!=\!\frac{-P_{\text{I}}
    \text{Tr}(\mathbf{q}_{\text{I}}^{H}\mathbf{W}\mathbf{q}_{\text{I}}\mathbf{U}_{\text{t}})
    \text{Tr}(\mathbf{q}_{\text{O}}^{H}\mathbf{W}\mathbf{q}_{\text{O}}\mathbf{U}_{\text{r}})\sigma^{2}}{
   \left(P_{\text{I}}\text{Tr}(\mathbf{q}_{E,\text{I}}\mathbf{q}_{E,\text{I}}^{H}\mathbf{U}_{\text{t}})+\sigma^{2}\right)
    \left(P_{\text{O}}\text{Tr}(\mathbf{q}_{\text{O}}^{H}\mathbf{W}\mathbf{q}_{\text{O}}\mathbf{U}_{\text{r}})+\sigma^{2}\right)^2}.
\end{gather}
It is clear that $\mathfrak{O}_{\text{I}}(P_{\text{I}},P_{\text{O}})$ is a monotonic decreasing function of $P_{\text{O}}$ because $\nabla_{P_{\text{O}}}\mathfrak{O}_{\text{I}}<0$ always holds. However, we can not directly judge the monotonicity of $\mathfrak{O}_{\text{I}}(P_{\text{I}},P_{\text{O}})$ with respect to $P_{\text{I}}$ from \eqref{28}. To further illustrate monotonicity of $\mathfrak{O}_{\text{I}}(P_{\text{I}},P_{\text{O}})$, we introduce following implicit conditions that if problem (24) is feasible, the optimized variables $\{\mathbf{W},\mathbf{U}_{\text{t}},\mathbf{U}_{\text{r}}\}$ in \textbf{Algorithm-1} will satisfy $\frac{P_{\text{I}}\text{Tr}(\mathbf{q}_{\text{I}}^{H}\mathbf{W}\mathbf{q}_{\text{I}}\mathbf{U}_{\text{t}})}
{P_{\text{O}}\text{Tr}(\mathbf{q}_{\text{O}}^{H}\mathbf{W}\mathbf{q}_{\text{O}}\mathbf{U}_{\text{r}})+\sigma^{2}}>
\frac{P_{\text{I}}\text{Tr}(\mathbf{q}_{E,\text{I}}\mathbf{q}_{E,\text{I}}^{H}\mathbf{U}_{\text{t}})}
{\sigma^2}$ and $\frac{P_{\text{O}}\text{Tr}(\mathbf{q}_{\text{O}}^{H}\mathbf{W}\mathbf{q}_{\text{O}}\mathbf{U}_{\text{r}})}
{\sigma^{2}}>\frac{P_{\text{O}}\text{Tr}(\mathbf{q}_{E,\text{O}}\mathbf{q}_{E,\text{O}}^{H}\mathbf{U}_{\text{r}})}{\sigma^2}$. This is because that if these two conditions do not hold, the minimum secrecy capacity returned by \textbf{Algorithm-1} is always less than $0$, which contradicts the constraint $\xi\geq0$. Therefore, by substituting $\frac{P_{\text{I}}\text{Tr}(\mathbf{q}_{\text{I}}^{H}\mathbf{W}\mathbf{q}_{\text{I}}\mathbf{U}_{\text{t}})}
{P_{\text{O}}\text{Tr}(\mathbf{q}_{\text{O}}^{H}\mathbf{W}\mathbf{q}_{\text{O}}\mathbf{U}_{\text{r}})+\sigma^{2}}>
\frac{P_{\text{I}}\text{Tr}(\mathbf{q}_{E,\text{I}}\mathbf{q}_{E,\text{I}}^{H}\mathbf{U}_{\text{t}})}
{\sigma^2}$ into \eqref{28}, we can derive $\nabla_{P_{\text{I}}}\mathfrak{O}_{\text{I}}>0$, which implies that $\mathfrak{O}_{\text{I}}(P_{\text{I}},P_{\text{O}})$ is a monotonic increasing function with respect to $P_{\text{I}}$. In the same way, we can also prove that function $\mathfrak{O}_{\text{O}}(P_{\text{O}})$ is a monotonic increasing function with respect to $P_{\text{O}}$.

\begin{proposition}
    The optimal transmit power policy under the case of $u_{\text{I}}=1,u_{\text{O}}=0$ is given by
    \begin{equation}\label{30}
    \begin{cases}
        P_{\text{O}}^{*}\!=\!\min\!\left\{\Big[\frac{2P_{\text{I},\text{max}}Z_{\text{E},\text{I}}Z_{\text{O}}\sigma^{2}-
        P_{\text{I},\text{max}}Z_{\text{I}}Z_{\text{E},\text{O}}\sigma^{2}+Z_{\text{O}}\sigma^{4}-Z_{\text{E},\text{O}}\sigma^{4}\pm\sqrt{\Delta_{\text{O}}}}
        {2(-P_{\text{I},\text{max}}Z_{\text{E},\text{I}}Z_{\text{O}}^{2}-Z_{\text{O}}^{2}\sigma^{2}+Z_{\text{E},\text{O}}Z_{\text{O}}\sigma^{2})}\Big]^{+},
        \frac{P_{\text{I},\text{max}}\text{Tr}(\mathbf{q}_{\text{I}}^{H}\mathbf{W}\mathbf{q}_{\text{I}}\mathbf{U}_{\text{t}})}
        {\text{Tr}(\mathbf{q}_{\text{O}}^{H}\mathbf{W}\mathbf{q}_{\text{O}}\mathbf{U}_{\text{r}})},
        P_{\text{O},\text{max}}\right\},\\
        P_{\text{I}}^{*}\!=\!P_{\text{I},\text{max}},
        \end{cases}
    \end{equation} where $\Delta_{\text{O}}\!=\!(2P_{\text{I},\text{max}}Z_{\text{E},\text{I}}Z_{\text{O}}\sigma^{2}\!-\!
    P_{\text{I},\text{max}}Z_{\text{I}}Z_{\text{E},\text{O}}\sigma^{2}\!+\!
    Z_{\text{O}}\sigma^{4}\!-\!Z_{\text{E},\text{O}}\sigma^{4})^{2}\!+\!
    4(P_{\text{I},\text{max}}Z_{\text{E},\text{I}}Z_{\text{O}}^{2}\!+\!Z_{\text{O}}^{2}\sigma^{2}\!-\!Z_{\text{E},\text{O}}Z_{\text{O}}\sigma^{2})
    (P_{\text{I},\text{max}}Z_{\text{I}}\sigma^{4}\!-\!P_{\text{I},\text{max}}Z_{\text{E},\text{I}}\sigma^{4})$,
     $Z_{\rho}\!=\!\text{Tr}(\mathbf{q}_{\rho}^{H}\mathbf{W}\mathbf{q}_{\rho}\mathbf{U}_{\text{t}/\text{r}})$ and
    $Z_{\text{E},\rho}\!=\!\text{Tr}(\mathbf{q}_{\text{E},\rho}\mathbf{q}_{\text{E},\rho}^{H}\mathbf{U}_{\text{t}/\text{r}})$ for $\rho\in\{\text{I},\text{O}\}$.
\end{proposition}
\begin{IEEEproof}
    With the monotonic property analyzed above, it is clear that the minimum secrecy rate reaches maximal value when $P_{\text{I}}=P_{\text{I},\text{max}}$. However, since $\mathfrak{O}_{\text{I}}(P_{\text{I}},P_{\text{O}})$ and $\mathfrak{O}_{\text{O}}(P_{\text{O}})$ are the monotonic decreasing and increasing functions of $P_{\text{O}}$, respectively, two possible situations probably occur when we increase $P_{\text{O}}$ from $0$ to $P_{\text{O},\text{max}}$: 1) there exists a value of $P_{\text{O}}^{*}$ between $0$ and $P_{\text{O},\text{max}}$, which satisfies $\mathfrak{O}_{\text{I}}(P_{\text{I}},P_{\text{O}}^{*})=\mathfrak{O}_{\text{O}}(P_{\text{O}}^{*})$; 2) there is no $P_{\text{O}}$ satisfying $\mathfrak{O}_{\text{I}}(P_{\text{I}},P_{\text{O}})=\mathfrak{O}_{\text{O}}(P_{\text{O}})$, i.e., $\mathfrak{O}_{\text{O}}(P_{\text{O}})<\mathfrak{O}_{\text{I}}(P_{\text{I}},P_{\text{O}})$ always holds. In the former case, the minimum secrecy rate reaches maximum when $\mathfrak{O}_{\text{I}}(P_{\text{I}},P_{\text{O}}^{*})=\mathfrak{O}_{\text{O}}(P_{\text{O}}^{*})$ holds, which can be simplified as $P_{\text{O}}^{*}=\left[\frac{2P_{\text{I},\text{max}}Z_{\text{E},\text{I}}Z_{\text{O}}\sigma^{2}-
        P_{\text{I},\text{max}}Z_{\text{I}}Z_{\text{E},\text{O}}\sigma^{2}+Z_{\text{O}}\sigma^{4}-Z_{\text{E},\text{O}}\sigma^{4}\pm\sqrt{\Delta_{\text{O}}}}
        {2(-P_{\text{I},\text{max}}Z_{\text{E},\text{I}}Z_{\text{O}}^{2}-Z_{\text{O}}^{2}\sigma^{2}+Z_{\text{E},\text{O}}Z_{\text{O}}\sigma^{2})}\right]^{+}$
        \footnote{Note that although we can obtain two candidate solutions via solving quadratic equation $\mathfrak{O}_{\text{I}}(P_{\text{I}},P_{\text{O}}^{*})=\mathfrak{O}_{\text{O}}(P_{\text{O}}^{*})$, there only exists a unique positive solution due to the monotonicity of $\mathfrak{O}_{\text{I}}(P_{\text{I}},P_{\text{O}})$ and $\mathfrak{O}_{\text{O}}(P_{\text{O}})$ with respect to $P_{\text{O}}$.}. In the latter case, the minimum secrecy rate equals to $\log_{2}(1+\mathfrak{O}_{\text{O}}(P_{\text{O}}))$, so the optimal $P_{\text{O}}^{*}$ is given by $P_{\text{O}}^{*}\!=\!P_{\text{O},\text{max}}$. Furthermore, considering the SIC decoding order constraint \eqref{25c}, $P_{\text{O}}$ is also limited by $\frac{P_{\text{I},\text{max}}\text{Tr}(\mathbf{q}_{\text{I}}^{H}\mathbf{W}\mathbf{q}_{\text{I}}\mathbf{U}_{\text{t}})}
        {\text{Tr}(\mathbf{q}_{\text{O}}^{H}\mathbf{W}\mathbf{q}_{\text{O}}\mathbf{U}_{\text{r}})}$. Accordingly, the closed-form optimal $P_{\text{O}}^{*}$ can be derived. This completes proof.
\end{IEEEproof}

By adopting Proposition 1, we can further derive the optimal $\{P_{\text{I}}^{*},P_{\text{O}}^{*}\}$ under the arbitrary SIC decoding order as follows.
\begin{equation}\label{31}
    \!\!\begin{cases}
        \!P_{\text{I}}^{*}\!=\!u_{\text{I}}P_{\text{I},\text{max}}\!+\!u_{\text{O}}\min\!\left\{\!\left[\frac{2P_{\text{O},\text{max}}Z_{\text{E},\text{O}}Z_{\text{I}}\sigma^{2}-
        P_{\text{O},\text{max}}Z_{\text{O}}Z_{\text{E},\text{I}}\sigma^{2}+Z_{\text{I}}\sigma^{4}-Z_{\text{E},\text{I}}\sigma^{4}\pm\sqrt{\Delta_{\text{I}}}}
        {2(-P_{\text{O},\text{max}}Z_{\text{E},\text{O}}Z_{\text{I}}^{2}-Z_{\text{I}}^{2}\sigma^{2}+Z_{\text{E},\text{I}}Z_{\text{I}}\sigma^{2})}\right]^{+}\!\!\!,\!
        \frac{P_{\text{O},\text{max}}\text{Tr}(\mathbf{q}_{\text{O}}^{H}\mathbf{W}\mathbf{q}_{\text{O}}\mathbf{U}_{\text{r}})}
        {\text{Tr}(\mathbf{q}_{\text{I}}^{H}\mathbf{W}\mathbf{q}_{\text{I}}\mathbf{U}_{\text{t}})},
        P_{\text{I},\text{max}}\!\right\},\\
        \!P_{\text{O}}^{*}\!=\!u_{\text{O}}P_{\text{O},\text{max}}\!+\!u_{\text{I}}\min\!\left\{\!\left[\frac{2P_{\text{I},\text{max}}Z_{\text{E},\text{I}}Z_{\text{O}}\sigma^{2}-
        P_{\text{I},\text{max}}Z_{\text{I}}Z_{\text{E},\text{O}}\sigma^{2}+Z_{\text{O}}\sigma^{4}-Z_{\text{E},\text{O}}\sigma^{4}\pm\sqrt{\Delta_{\text{O}}}}
        {2(-P_{\text{I},\text{max}}Z_{\text{E},\text{I}}Z_{\text{O}}^{2}-Z_{\text{O}}^{2}\sigma^{2}+Z_{\text{E},\text{O}}Z_{\text{O}}\sigma^{2})}\right]^{+}\!\!\!,\!
        \frac{P_{\text{I},\text{max}}\text{Tr}(\mathbf{q}_{\text{I}}^{H}\mathbf{W}\mathbf{q}_{\text{I}}\mathbf{U}_{\text{t}})}
        {\text{Tr}(\mathbf{q}_{\text{O}}^{H}\mathbf{W}\mathbf{q}_{\text{O}}\mathbf{U}_{\text{r}})},
        P_{\text{O},\text{max}}\!\right\},
    \!\!\end{cases}
\end{equation}
where $\Delta_{\text{I}}=(2P_{\text{O},\text{max}}Z_{\text{E},\text{O}}Z_{\text{I}}\sigma^{2}\!-\!
    P_{\text{O},\text{max}}Z_{\text{O}}Z_{\text{E},\text{I}}\sigma^{2}\!+\!
    Z_{\text{I}}\sigma^{4}\!-\!Z_{\text{E},\text{I}}\sigma^{4})^{2}\!+\!
    4(P_{\text{O},\text{max}}Z_{\text{E},\text{O}}Z_{\text{I}}^{2}\!+\!Z_{\text{I}}^{2}\sigma^{2}\!-\!Z_{\text{E},\text{I}}Z_{\text{I}}\sigma^{2})
    (P_{\text{O},\text{max}}Z_{\text{O}}\sigma^{4}\!-\!P_{\text{O},\text{max}}Z_{\text{E},\text{O}}\sigma^{4})$.
\vspace{-4mm}
\subsection{Overall Algorithm}
The overall algorithm is summarized in \textbf{Algorithm-2}, where $R_{\text{s},\text{min}}(n)$ is introduced to record the minimum security capacity of each alternate iteration, while $\varepsilon_{\text{th}}$ is the pre-defined convergence accuracy of AHB algorithm.

\begin{table}[h]
    \centering
    \begin{tabular}{p{450pt}}
    \toprule
    \textbf{Algorithm-2:} Proposed AHB Algorithm for Minimum Secrecy Capacity Maximization\\
    \midrule
    1: \textbf{Initialization}: Initialize the $P_{\text{I}}(n)$, $P_{\text{O}}(n)$ and $R_{\text{s},\text{min}}(n)=+\infty$ with $n=1$;\\
    \hangafter 1 
    \hangindent 2em 
    2: \textbf{Repeat }\\
    3:          \quad Perform \textbf{Algorithm-1} to obtain the stationary point solutions $\{\mathbf{W}(n),\mathbf{U}_{\text{t}}(n),\mathbf{U}_{\text{r}}(n)\}$ with the fixed $\{P_{\text{I}}(n),P_{\text{O}}(n),$\\
    \qquad $u_{\text{I}},u_{\text{O}}\}$;\\
    4:          \quad Calculate the optimal solutions $\{P_{\text{I}}^{*}(n+1),P_{\text{O}}^{*}(n+1)\}$ according to \eqref{31} with the fixed  $\{\mathbf{W}(n),\mathbf{U}_{\text{t}}(n),\mathbf{U}_{\text{r}}(n),$\\
    \qquad $u_{\text{I}},u_{\text{O}}\}$;\\
    5:          \quad Update $R_{\text{s},\text{min}}(n+1)$ and set $n = n+1$;\\
    6:  \textbf{Until} $|R_{\text{s},\text{min}}(n)-R_{\text{s},\text{min}}(n-1)|\leq\varepsilon_{\text{th}}$.\\
    \bottomrule
    \end{tabular}
\end{table}

\begin{proposition}
    The proposed AHB algorithm is guaranteed to converge to the suboptimal solution over the non-decreasing iterations.
\end{proposition}
\begin{IEEEproof}
    See Appendix B.
\end{IEEEproof}

Since the optimal transmit power are updated by the closed-form expression, the main computational complexity of \textbf{Algorithm-2} relies on the complexity of \textbf{Algorithm-1}. The complexity to solve (24) via the standard interior-point method is given by $\mathcal{O}\Big(l_{\text{A}}l_{\text{outer}}l_{\text{inner}}\sqrt{2N+M+8}\\ p(p^{2}+p(2N^{2}+M^{2})+2N^{3}+M^{3}+4pq^{2})\Big)$ \cite{A.Ben-Tal_complexity}, where $p=2N^{2}+M^{2}+12$ denotes the number of variables, $q=N^{2}+1$ denotes the dimension of the second-order cone (SOC) constraint \cite{Miguel_SOC_demension}, while $l_{\text{A}}$, $l_{\text{outer}}$ and $l_{\text{inner}}$ denote the numbers of the alternating iteration, the outer layer iteration, and the inner layer iteration.

\section{Proposed Solution for Statistical Eavesdropping CSI}
In this section, we extend the AHB algorithm proposed in the previous section to solve problem (10). Specifically, we first derive the exact SOP expression of IU/OU, and decompose the problem (10) into two subproblems. For the joint beamforming optimization subproblem, the modified two-layer iterative algorithm is developed to optimize reflection/transmission coefficients and receiving beamforming jointly. While for the power optimization subproblem, we derive the optimal transmit power policy in the closed-form expression.

\vspace{-4mm}
\subsection{Exact SOP Expression of IU and OU}
Let $\xi_{\text{E},\text{I}}=\frac{\mathbf{h}_{\text{E},\text{S}}^{H}\bm{\Theta}^{\text{t}}\mathbf{h}_{\text{I},\text{S}}}
{L_{\text{E},\text{S}}L_{\text{I},\text{S}}}=\sum_{n=1}^{N}\widetilde{h}_{\text{E},\text{S}_{n}}^{H}
\sqrt{\beta^{\text{t}}_{n}}e^{j\theta^{\text{t}}_{n}}\widetilde{h}_{\text{I},\text{S}_{n}}$ and $\xi_{\text{E},\text{O}}=\frac{\mathbf{h}_{\text{E},\text{S}}^{H}\bm{\Theta}^{\text{r}}\mathbf{h}_{\text{O},\text{S}}}{L_{\text{E},\text{S}}L_{\text{O},\text{S}}}
=\sum_{n=1}^{N}\widetilde{h}_{\text{E},\text{S}_{n}}^{H}
\sqrt{\beta^{\text{r}}_{n}}e^{j\theta^{\text{r}}_{n}}\\\widetilde{h}_{\text{O},\text{S}_{n}}$, where $\widetilde{h}_{\text{E},\text{S}_{n}}, \widetilde{h}_{\text{I},\text{S}_{n}},\widetilde{h}_{\text{O},\text{S}_{n}}\sim \mathcal{CN}(0,1)$ denote the small fading coefficients, while $L_{\text{E},\text{S}}$, $L_{\text{I},\text{S}}$ and $L_{\text{O},\text{S}}$ denote the large scale path loss factors. To reveal the distribution of random variables $\xi_{\text{E},\text{I}}$ and $\xi_{\text{E},\text{O}}$, we first derive the exact probability density function (PDF) of $\xi_{\text{E},\text{I}}^{n}$ ($1\leq n\leq N$), which is defined as $\xi_{\text{E},\text{I}}^{n}=\widetilde{h}_{\text{E},\text{S}_{n}}^{H}
\sqrt{\beta^{\text{t}}_{n}}e^{j\theta^{\text{t}}_{n}}\widetilde{h}_{\text{I},\text{S}_{n}}$. Let $\widetilde{h}_{\text{E},\text{S}_{n}}=x_{n}-iy_{n}$, $\widetilde{h}_{\text{I},\text{S}_{n}}=p_{n}+iq_{n}$ and $\sqrt{\beta^{\text{t}}_{n}}e^{j\theta^{\text{t}}_{n}}=\sqrt{\beta^{\text{t}}_{n}}(\text{cos}\theta^{\text{t}}_{n}+i\text{sin}\theta^{\text{t}}_{n})$, where the real random variables $x_{n}$, $y_{n}$, $p_{n}$ and $q_{n}$ follow the independent and identically distributed (i.i.d.) Gaussian distribution with zero mean and $\frac{1}{2}$ variance. Thus, we can rewrite $\xi_{\text{E},\text{I}}^{n}$ as
\begin{equation}
\label{32}
\xi_{\text{E},\text{I}}^{n}=\sqrt{\beta^{\text{t}}_{n}}\Big(\underbrace{x_{n}\widetilde{p}_{n}-y_{n}\widetilde{q}_{n}}_{a_{n}}+
i(\underbrace{x_{n}\widetilde{q}_{n}+y_{n}\widetilde{p}_{n}}_{b_{n}})\Big),
\end{equation}
where $\widetilde{p}_{n}=\text{cos}\theta^{\text{t}}_{n}p_{n}-\text{sin}\theta^{\text{t}}_{n}q_{n}$ and $\widetilde{q}_{n}=\text{sin}\theta^{\text{t}}_{n}p_{n}+\text{cos}\theta^{\text{t}}_{n}q_{n}$. By treating $\theta^{\text{t}}_{n}$, $p_{n}$ and $q_{n}$ as the constant coefficients, $a_{n}$ and $b_{n}$ can be regarded as the sums of the i.i.d. Gaussian variables $x_{n}$ and $y_{n}$. Thus, $a_{n}$ and $b_{n}$ follow the Gaussian distribution, i.e., $a_{n},b_{n}\sim \mathcal{CN}(0,\frac{\widetilde{p}_{n}^{2}+\widetilde{q}_{n}^{2}}{2})$ \cite{Z.Ding_IRS_NOMA_2}.

While for any linear combination of $a_{n}$ and $b_{n}$, we can rewrite it as
\begin{equation}
\label{34}
\psi_{1}a_{n}+\psi_{2}b_{n}=x_{n}\begin{bmatrix}\widetilde{p}_{n},& \widetilde{q}_{n} \end{bmatrix}\begin{bmatrix}\psi_{1}\\ \psi_{2}\end{bmatrix}-y_{n}\begin{bmatrix}\widetilde{p}_{n},& \widetilde{q}_{n}\end{bmatrix}\begin{bmatrix}-\psi_{2}\\ \psi_{1}\end{bmatrix}.
\end{equation}
Note $\widetilde{p}_{n}$ and $\widetilde{q}_{n}$ are also Gaussian variables as they are the sums of i.i.d. Gaussian variables $p_{n}$ and $q_{n}$. Obviously, $[\widetilde{p}_{n}, \widetilde{q}_{n}][\psi_{1}, \psi_{2}]^{T}$ and $[\widetilde{p}_{n}, \widetilde{q}_{n}][-\psi_{2},\psi_{1}]^{T}$ consist of two i.i.d. Gaussian variables $\widetilde{p}_{n}$ and $\widetilde{q}_{n}$ with orthogonal weight coefficient vectors $[\psi_{1}, \psi_{2}]^{T}$ and $[-\psi_{2},\psi_{1}]^{T}$, so they follows i.i.d. Gaussian distribution. Since the sum of i.i.d. Gaussian variables is also Gaussian distributed, $\psi_{1}a_{n}+\psi_{2}b_{n}$ follows Gaussian distribution, which indicates that $a_{n}$ and $b_{n}$ follow two-dimensional Gaussian distribution. Moreover, with the independence of $x_{n}$, $y_{n}$, $\widetilde{p}_{n}$ and $\widetilde{q}_{n}$, the correlation between $a_{n}$ and $b_{n}$ can be expressed as
\begin{align}
\label{33}\nonumber
\mathbb{E}\{a_{n}b_{n}\}&=\mathbb{E}\{(x_{n}\widetilde{p}_{n}-y_{n}\widetilde{q}_{n})(x_{n}\widetilde{q}_{n}+y_{n}\widetilde{p}_{n})\},\\ \nonumber
&=\mathbb{E}\{x_{n}^{2}\}\mathbb{E}\{\widetilde{p}_{n}\}\mathbb{E}\{\widetilde{q}_{n}\}\!+\!\mathbb{E}\{x_{n}\}\mathbb{E}\{y_{n}\}\mathbb{E}\{\widetilde{p}_{n}^{2}\}\!-\!
\mathbb{E}\{x_{n}\}\mathbb{E}\{y_{n}\}\mathbb{E}\{\widetilde{q}_{n}^{2}\}\!- \! \mathbb{E}\{y_{n}^{2}\}\mathbb{E}\{\widetilde{p}_{n}\}\mathbb{E}\{\widetilde{q}_{n}\},\\
&=0,
\end{align}
which proves the independence of $a_{n}$ and $b_{n}$. Up to now, we have proved that $a_{n}$ and $b_{n}$ are the i.i.d. Gaussian variables, which also follow joint Gaussian distribution and independent (also uncorrelated), so the complex variable $\xi_{\text{E},\text{I}}^{n}$ follows complex Gaussian distribution, i.e., $\xi_{\text{E},\text{I}}^{n}\sim \mathcal{CN}(0,\beta^{\text{t}}_{n}(\widetilde{p}_{n}^{2}+\widetilde{q}_{n}^{2}))$, where $\widetilde{p}_{n}^{2}+\widetilde{q}_{n}^{2}=p_{n}^{2}+q_{n}^{2}=|\widetilde{h}_{\text{I},\text{S}_{n}}|^{2}$. To proceed, by summing the real parts and imaginary parts of $\xi_{\text{E},\text{I}}^{n}$ and $\xi_{\text{E},\text{I}}^{m}$ ($n\neq m$) separately, we have that $\xi_{\text{E},\text{I}}\sim \mathcal{CN}(0,\sum_{n=1}^{N}\beta^{\text{t}}_{n}|\widetilde{h}_{\text{I},\text{S}_{n}}|^{2})$. Note that this result is also applicable to $\xi_{\text{E},\text{O}}$, i.e., $\xi_{\text{E},\text{O}}\sim \mathcal{CN}(0,\sum_{n=1}^{N}\beta^{\text{r}}_{n}|\widetilde{h}_{\text{O},\text{S}_{n}}|^{2})$. Then, the PDF of $|\xi_{\text{E},\rho}|^{2}$ is an exponential distribution function with parameter $\frac{1}{\sum_{n=1}^{N}\beta^{\text{t}/\text{r}}_{n}|\widetilde{h}_{\rho,\text{S}_{n}}|^{2}}$, $\rho\in\{\text{I},\text{O}\}$.

Accordingly, we can derive the exact SOP as
\begin{align}
\label{35}\nonumber
P_{\text{SOP},\rho}&=\mathbb{P}\left(|\xi_{\text{E},\rho}|^{2}>\frac{(2^{R_{\text{c},\rho}-R_{\text{s},\rho}}-1)\sigma^{2}}{P_{\rho}L_{\text{E},\text{S}}^{2}L_{\rho,\text{S}}^{2}}\right),\\ \nonumber
&=\int_{\frac{(2^{R_{\text{c},\rho}-R_{\text{s},\rho}}-1)\sigma^{2}}{P_{\rho}L_{\text{E},\text{S}}^{2}L_{\rho,\text{S}}^{2}}}^{\infty}\frac{1}{\sum_{n=1}^{N}\beta^{\text{t}/\text{r}}_{n}|\widetilde{h}_{\rho,\text{S}_{n}}|^{2}}
e^{-\frac{|\xi_{\text{E},\rho}|^{2}}{\sum_{n=1}^{N}\beta^{\text{t}/\text{r}}_{n}|\widetilde{h}_{\rho,\text{S}_{n}}|^{2}}}d|\xi_{\text{E},\rho}|^{2},\\
&=e^{-\frac{(2^{R_{\text{c},\rho}-R_{\text{s},\rho}}-1)\sigma^{2}}{P_{\rho}L_{\text{E},\text{S}}^{2}L_{\rho,\text{S}}^{2}\sum_{n=1}^{N}\beta^{\text{t}/\text{r}}_{n}|\widetilde{h}_{\rho,\text{S}_{n}}|^{2}}}, \quad \rho\in\{\text{I},\text{O}\}.
\end{align}

\begin{figure}[t]
  \centering
  \includegraphics[scale = 0.45]{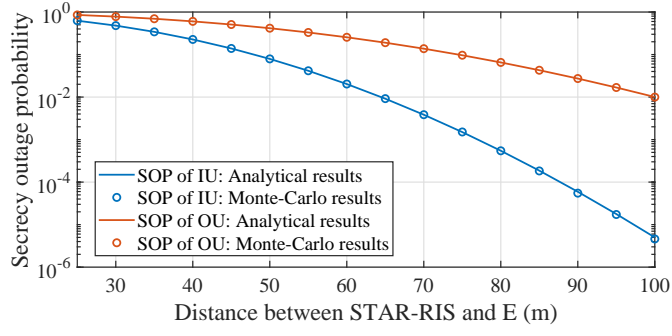}
  \caption{Tightness of the analytical SOP expression with $M=8$, $N=24$ and $R_{\text{c},\rho}-R_{\text{s},\rho}=1$bps/Hz.} \vspace{-8mm}
  \label{Fig.2}
\end{figure}

\textit{Tightness of the analytical SOP expression}: The analytical/numerical SOP of IU and OU versus the distance between the STAR-RIS and E is illustrated in Fig. \ref{Fig.2}, where the random phase shifts and amplitudes are considered, and the other simulation parameters are the same as Table I. As can be observed, 1) the analytical expression of SOP in \eqref{35} for both two users are accurate; and 2) the secrecy performance of IU behaves worse than that of OU, which is because IU is closer to E and suffers stronger eavesdropping links.



\vspace{-2mm}
\subsection{Joint Beamforming and Power Optimization}
\textit{(1) Optimizing beamforming with fixed $\{P_{\text{I}},P_{\text{O}}\}$}: Based on the derivations above, problem (10) is reduced to
\vspace{-5mm}
\begin{subequations}
\begin{align}
\label{36a} &\min\limits_{\mathbf{U}_{\text{t}},\mathbf{U}_{\text{r},\varsigma_{\rho}},\mathbf{W}}\quad \max\limits_{\rho\in\{\text{I},\text{O}\}}\ \varsigma_{\rho},\\
\label{36b}&\quad\text{s.t.} \quad\ \  \eqref{11b}-\eqref{11f},\\
\label{36c}&\quad\quad\quad\,\, \gamma_{\rho}\geq 2^{R_{\text{c},\rho}}-1,\quad \rho\in\{\text{I},\text{O}\},
\end{align}
\end{subequations}
where $\varsigma_{\rho}$ is a weighted linear combination of $\mathbf{U}_{\text{t}/\text{r}}(n,n)$, i.e., $\frac{P_{\rho}L_{\text{E},\text{S}}^{2}L_{\rho,\text{S}}^{2}}{2^{R_{\text{c},\rho}-R_{\text{s},\rho}}-1}\sum_{n=1}^{N}\mathbf{U}_{\text{t}/\text{r}}(n,n)|
\widetilde{h}_{\rho,\text{S}_{n}}|^{2}$. Similarly, we introduce the slack variables $\mathfrak{T}_{\rho,\text{t}/\text{r}}^{\text{lower}}$ and $\mathfrak{T}_{\rho,\text{t}/\text{r}}^{\text{upper}}$ satisfying \eqref{13} and \eqref{14}, and thus the constraint \eqref{11b} can be transformed into \eqref{15}, while \eqref{36c} can be rewritten as the following linear form
\begin{align}
\label{37}\begin{cases}
P_{\text{I}}\mathfrak{T}_{\text{I},\text{t}}^{\text{lower}}\geq
(P_{\text{O}}\mathfrak{T}_{\text{O},\text{r}}^{\text{upper}}\!+\!\sigma^2)(2^{R_{\text{c},\text{I}}-R_{\text{s},\text{I}}}\!-\!1),\ \  \frac{P_{\text{O}}\mathfrak{T}_{\text{O},\text{r}}^{\text{lower}}}
{\sigma^2}\geq 2^{R_{\text{c},\text{O}}-R_{\text{s},\text{O}}}\!-\!1, \ \text{if $u_{\text{I}}=1$, $u_{\text{O}}=0$,}\\
\frac{P_{\text{I}}\mathfrak{T}_{\text{I},\text{t}}^{\text{lower}}}
{\sigma^2}\geq 2^{R_{\text{c},\text{I}}-R_{\text{s},\text{I}}}\!-\!1,\ \  P_{\text{O}}\mathfrak{T}_{\text{O},\text{r}}^{\text{lower}}\geq
(P_{\text{I}}\mathfrak{T}_{\text{I},\text{t}}^{\text{upper}}+\sigma^2)(2^{R_{\text{c},\text{O}}-R_{\text{s},\text{O}}}\!-\!1),\  \text{if $u_{\text{I}}=0$, $u_{\text{O}}=1$.}
\end{cases}
\end{align}
To convexify the rank-one constraints \eqref{11f}, we can rewrite $\text{rank}(\mathbf{U}_{\text{t}/\text{r}})=1$ as \eqref{23}. As for $\text{rank}(\mathbf{W})=1$, we can observe that Lagrangian functions with respect to $\mathbf{W}$ of problems (24) and (36) possess the same form, which implies that the results in Lemma 2 is also applicable to problem (36). Therefore, problem (36) is reformulated  as
\begin{subequations}
\begin{align}
\label{38a} &\min\limits_{\mathbf{U}_{\text{t}},\mathbf{U}_{\text{r}},\mathbf{W},
\atop
\mathfrak{T}_{\rho,\text{t}/\text{r}}^{\text{upper}},\mathfrak{T}_{\rho,\text{t}/\text{r}}^{\text{lower}},
\varsigma_{\rho},\varrho_{\text{t}/\text{r}}}\quad \max\limits_{\rho\in\{\text{I},\text{O}\}}\ \varsigma_{\rho}
-\tau(\varrho_{\text{t}}+\varrho_{\text{r}}),\\
\label{38b}&\quad\text{s.t.} \quad\ \  \eqref{11b}-\eqref{11e},\eqref{13}-\eqref{15},\eqref{23},\eqref{37}.
\end{align}
\end{subequations}
Note that problem (38) is a convex optimization problem with rank-one penalty terms and can be solved similarly as \textbf{Algorithm-1}, so we omit its algorithm for brevity.

\textit{(2) Optimizing transmit power with fixed $\{\mathbf{U}_{\text{t}},\mathbf{U}_{\text{r}},\mathbf{W}\}$}: With given $\{\mathbf{U}_{\text{t}},\mathbf{U}_{\text{r}},\mathbf{W}\}$, problem (10) is reduced into
\begin{subequations}
\begin{align}
\label{39a} &\min\limits_{P_{\text{I}},P_{\text{O}},\varsigma_{\rho}}\quad \max\limits_{\rho\in\{\text{I},\text{O}\}}\ \varsigma_{\rho},\\
\label{39b}&\quad\text{s.t.} \quad\ \  \eqref{8b},\eqref{10c},\eqref{11b}.
\end{align}
\end{subequations}
By leveraging Proposition 2, we can derive the optimal transmit power policy in the closed-form expression.
\begin{proposition}
    If problem (39) is feasible, the optimal transmit power $\{P_{\text{I}}^{*},P_{\text{O}}^{*}\}$ under the arbitrary decoding order is given by
    \begin{equation}\label{40}
    \begin{cases}
    \!P_{\text{I}}^{*}\!=\!
    u_{\text{I}}\max\Big\{\!\frac{(P_{\text{O}}^{*}\text{Tr}(\mathbf{q}_{\text{O}}^{H}\mathbf{W}\mathbf{q}_{\text{O}}\mathbf{U}_{\text{r}})
    \!+\!\sigma^2)(2^{R_{\text{c},\text{I}}-R_{\text{s},\text{I}}}\!-\!1)}
    {\text{Tr}(\mathbf{q}_{\text{I}}^{H}\mathbf{W}\mathbf{q}_{\text{I}}\mathbf{U}_{\text{t}})},
    \frac{P_{\text{O}}^{*}\text{Tr}(\mathbf{q}_{\text{O}}^{H}\mathbf{W}\mathbf{q}_{\text{O}}\mathbf{U}_{\text{r}})}
    {\text{Tr}(\mathbf{q}_{\text{I}}^{H}\mathbf{W}\mathbf{q}_{\text{I}}\mathbf{U}_{\text{t}})}\!\Big\}+u_{\text{O}}\frac{\sigma^2(2^{R_{\text{c},\text{I}}-R_{\text{s},\text{I}}}\!-\!1)}
    {\text{Tr}(\mathbf{q}_{\text{I}}^{H}\mathbf{W}\mathbf{q}_{\text{I}}\mathbf{U}_{\text{t}})},\\
    \!P_{\text{O}}^{*}\!=\!u_{\text{O}}\max\Big\{\!
    \frac{(P_{\text{I}}^{*}\text{Tr}(\mathbf{q}_{\text{I}}^{H}\mathbf{W}\mathbf{q}_{\text{I}}\mathbf{U}_{\text{t}})
    +\sigma^2)(2^{R_{\text{c},\text{O}}-R_{\text{s},\text{O}}}\!-\!1)}
    {\text{Tr}(\mathbf{q}_{\text{O}}^{H}\mathbf{W}\mathbf{q}_{\text{O}}\mathbf{U}_{\text{r}})},
    \frac{P_{\text{I}}^{*}\text{Tr}(\mathbf{q}_{\text{I}}^{H}\mathbf{W}\mathbf{q}_{\text{I}}\mathbf{U}_{\text{t}})}
    {\text{Tr}(\mathbf{q}_{\text{O}}^{H}\mathbf{W}\mathbf{q}_{\text{O}}\mathbf{U}_{\text{r}})}\!\Big\}+u_{\text{I}}\frac{\sigma^2(2^{R_{\text{c},\text{O}}-R_{\text{s},\text{O}}}\!-\!1)}
    {\text{Tr}(\mathbf{q}_{\text{O}}^{H}\mathbf{W}\mathbf{q}_{\text{O}}\mathbf{U}_{\text{r}})}
    , \!\!\end{cases}
    \end{equation}
\end{proposition}
\begin{IEEEproof}
    The objective function of problem (39) is $\max\{\ \frac{P_{\text{I}}L_{\text{E},\text{S}}^{2}L_{\text{I},\text{S}}^{2}}{2^{R_{\text{c},\text{I}}-R_{\text{s},\text{I}}}-1}
    \sum_{n=1}^{N}\mathbf{U}_{\text{t}}(n,n)|\widetilde{h}_{\text{I},\text{S}_{n}}|^{2},\\
    \frac{P_{\text{O}}L_{\text{E},\text{S}}^{2}L_{\text{O},\text{S}}^{2}}{2^{R_{\text{c},\text{O}}-R_{\text{s},\text{O}}}-1}
    \sum_{n=1}^{N}\mathbf{U}_{\text{r}}(n,n)|\widetilde{h}_{\text{O},\text{S}_{n}}|^{2}\}$, which is a monotonic increasing function with respect to $P_{\text{I}}$ and $P_{\text{O}}$. Thus, the minimum objective value is obtained only when constraints \eqref{10c} is active while the SIC decoding order constraint \eqref{11b} is also satisfied. Accordingly, the optimal transmit power policy can be derived as the results in \eqref{40}. This completes the proof.
\end{IEEEproof}
\vspace{-6mm}
\subsection{Overall Algorithm}
The overall algorithm is summarized in \textbf{Algorithm-3}, where $P_{\text{SOP},\text{max}}(n)$ is introduced to record the maximum SOP of each alternate iteration, while $\varepsilon_{\text{th}}$ is the pre-defined accuracy of the extended AHB algorithm.
\begin{table}[h]
    \centering
    \begin{tabular}{p{450pt}}
    \toprule
    \textbf{Algorithm-3:} Extended AHB Algorithm for Maximum SOP Minimization\\
    \midrule
    1: \textbf{Initialization}: Initialize the $P_{\text{I}}(n)$, $P_{\text{O}}(n)$ and $P_{\text{SOP},\text{max}}(n)=+\infty$ with $n=1$;\\
    \hangafter 1 
    \hangindent 2em 
    2: \textbf{Repeat }\\
    3:          \quad Solve problem (38) to obtain stationary point solutions $\{\mathbf{W}(n),\mathbf{U}_{\text{t}}(n),\mathbf{U}_{\text{r}}(n)\}$ with the fixed $\{P_{\text{I}}(n),P_{\text{O}}(n),u_{\text{I}},u_{\text{O}}\}$;\\
    4:          \quad Calculate the optimal solutions $\{P_{\text{I}}^{*}(n+1),\!P_{\text{O}}^{*}(n+1)\}$ according to \eqref{40} with the fixed $\{\mathbf{W}(n),\!\mathbf{U}_{\text{t}}(n),\! \mathbf{U}_{\text{r}}(n),u_{\text{I}},u_{\text{O}}\}$;\\
    5:          \quad Update $P_{\text{SOP},\text{max}}(n+1)$ and set $n = n+1$;\\
    6:  \textbf{Until} $|P_{\text{SOP},\text{max}}(n)-P_{\text{SOP},\text{max}}(n-1)|\leq\varepsilon_{\text{th}}$;\\
    \bottomrule
    \end{tabular}
\end{table}

By exploiting Proposition 2, it is easy to prove that the proposed extended AHB algorithm in \textbf{Algorithm-3} is guaranteed to converge to the suboptimal solution over the non-increasing iterations. Moreover, the computational complexity of the overall algorithm is given by $\mathcal{O}\Big(l_{\text{A}}l_{\text{outer}}l_{\text{inner}}\\\sqrt{2N+M+8} p(p^{2}+p(2N^{2}+M^{2})+2N^{3}+M^{3}+4pq^{2})\Big)$, where $l_{\text{A}}$, $l_{\text{outer}}$ and $l_{\text{inner}}$ respectively denote the numbers of the alternating iteration, the outer layer iteration, and the inner layer iteration, $p=2N^{2}+M^{2}+8$ denotes the number of the optimization variables and $q=N^{2}+1$ denotes the size of the SOC constraint.

\begin{table}[t]
	\centering  
    \vspace{-4mm}
	\caption{System parameters}  
	\label{table1}  
    \vspace{-6mm}
	\begin{tabular}{|c|c|}
		\hline  
		The path loss at the reference distance of 1 meter & $L_{0}=-30\text{dB}$ \\  
		\hline
		The path-loss exponent of the STAR-RIS-BS channel & $\alpha_{\text{B},\text{S}}=2.2$ \\
		\hline
        The path-loss exponents of the STAR-RIS-users channels & $\alpha_{\text{I},\text{S}}=\alpha_{\text{O},\text{S}}=\alpha_{\text{E},\text{S}}=2.5$\\
		\hline
        The noise power at receivers &  $\sigma^{2}=-115$dBm\\
		\hline \tabincell{c}{convergence accuracy }
         & $\xi_{\text{th}}=10^{-3}$, $\varrho_{\text{th}}=10^{-3}$ and $\varepsilon_{\text{th}}=10^{-4}$ \\
		\hline
	\end{tabular}
\end{table}
\vspace{-4mm}
\begin{figure}[t]
  \centering
  \includegraphics[scale = 0.6]{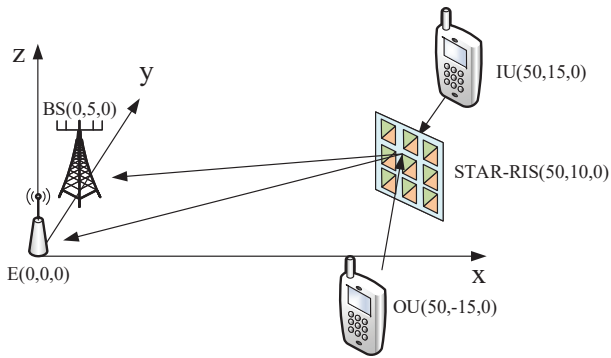}\vspace{-4mm}
  \caption{Simulation setup of the STAR-RIS assisted uplink NOMA secrecy communication.}\vspace{-8mm}
  \label{Fig.3}
\end{figure}

\section{Simulation Results}

In this section, we provide the numerical results to demonstrate the effectiveness of the proposed algorithms. As shown in Fig. 3, a three-dimensional coordinate network setup is considered, where the BS is located at $(0,5,0)$ meter (m), the STAR-RIS is located at $(50,10,0)$ m, the E is located at $(0,0,0)$ m, while the IU and OU are located at $(50,15,0)$ m and $(50,-15,0)$ m, respectively. The large-scale path loss model is given by $L=L_{0}(d)^{-\alpha}$, in which $L_{0}$ denote the path loss at the reference distance $1$m, while $d$ and $\alpha$ denotes the distance and the path loss exponent between the corresponding transceiver. The main simulation parameters are listed in Table I, and the other parameters are listed in the caption of each simulation figure. Furthermore, each result is the average over $100$ independent Monte-Carlo experiments.

\begin{figure}[t]
\centering 
\begin{minipage}[b]{0.45\textwidth} 
\centering 
\includegraphics[width=1\textwidth]{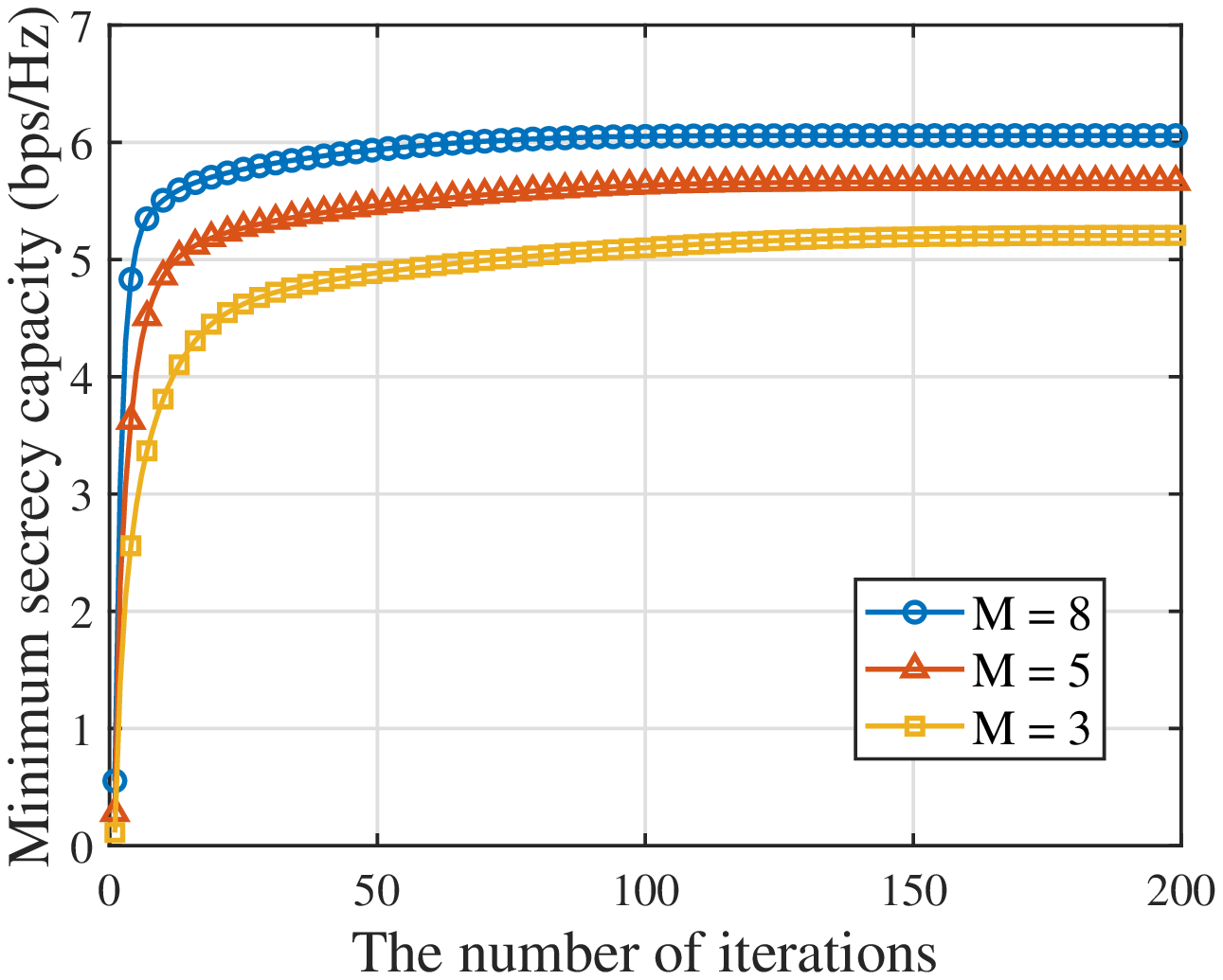}
\caption{The convergence of \textbf{Algorithm-2} with $N=20$ and $P_{\text{I},\text{max}}=P_{\text{O},\text{max}}=15$dBm.}
\label{Fig.4}
\end{minipage}\qquad
\begin{minipage}[b]{0.45\textwidth} 
\centering 
\includegraphics[width=1\textwidth]{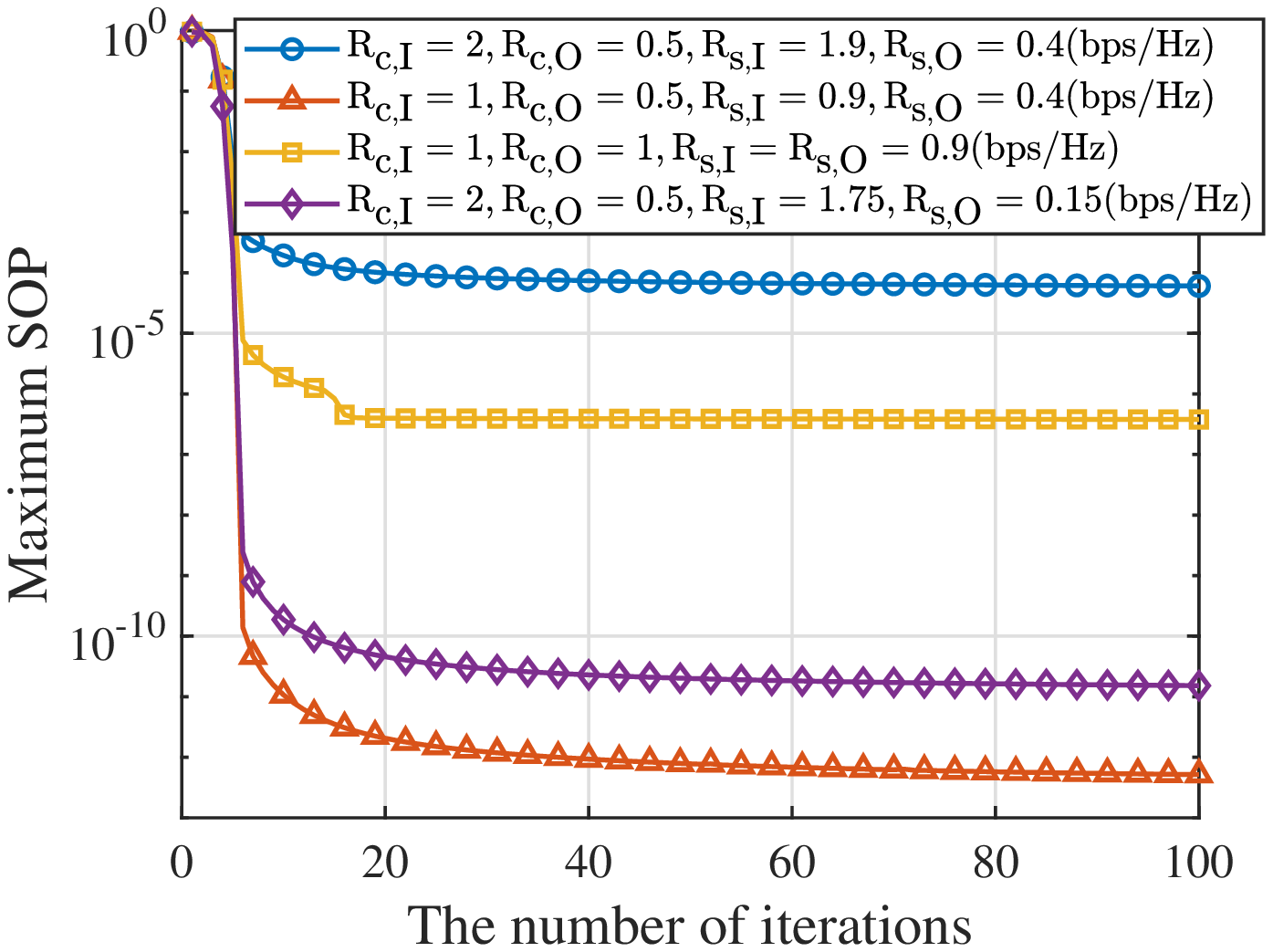} 
\caption{The convergence of \textbf{Algorithm-3} with $N=20$, $M=8$ and $P_{\text{I},\text{max}}=P_{\text{O},\text{max}}=15$dBm.}
\label{Fig.5}
\end{minipage} \vspace{-8mm}
\end{figure}

The convergence performances of Algorithm-2 and Algorithm-3 are illustrated in Fig. \ref{Fig.4} and Fig. \ref{Fig.5}, where both two algorithms are guaranteed to converge to the stable solutions over the iterations. In Fig. \ref{Fig.4}, it can be observed that a larger $M$ can achieve a larger minimum secrecy capacity and converge faster. An intuitive explanation to this interesting phenomenon is that: 1) increasing $M$ increases the spatial degree-of-freedom (DoF), which provides a more flexible receive beamforming design and achieves better secrecy performance; and 2) although a larger $M$ increases the computational complexity of the problem, it improves the optimization degree of each iteration, which helps reduce the number of iterations. It is also observed from Fig. \ref{Fig.5} that decreasing the codeword rate $R_{\text{c},\rho}$ and the secrecy rate $R_{\text{s},\rho}$ is conducive to reduce the SOP of the considered network. It is since that decreasing $R_{\text{c},\rho}$ decreases the QoS requirements of IU and OU, which limits the transmit power at the IU/OU, so as to degrade the reception ability of E. Also, a smaller $R_{\text{s},\rho}$ implies that more redundant rate is utilized to confuse E, which thus enhances the secrecy performance of the network.

\begin{figure}[t]
\centering 
\begin{minipage}[b]{0.45\textwidth} 
\centering 
\includegraphics[width=1\textwidth]{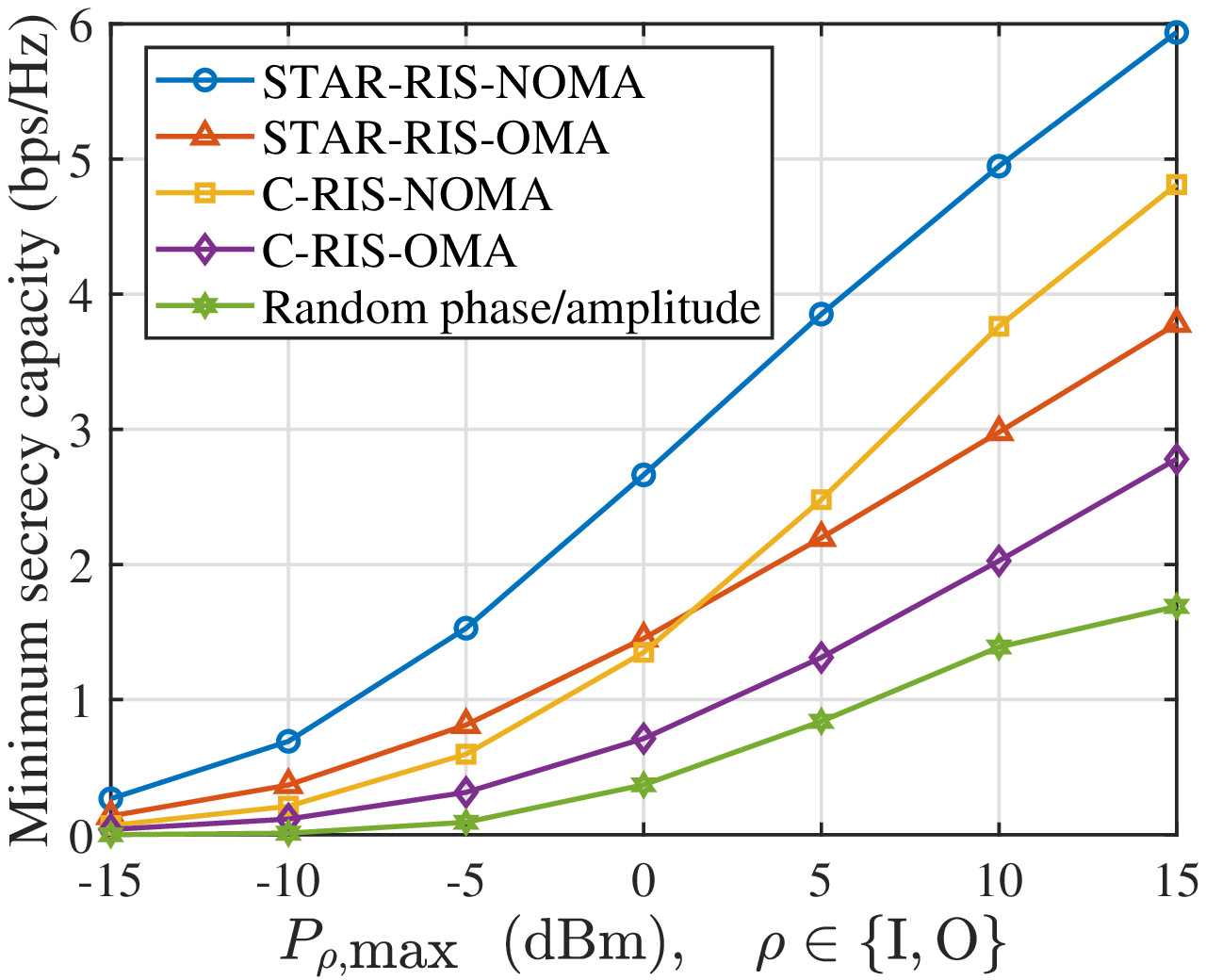}
\caption{The minimum secrecy capacity among IU and OU versus the maximal transmit power of IU/OU with $N=20$ and $M=8$.}
\label{Fig.6}
\end{minipage}\qquad
\begin{minipage}[b]{0.45\textwidth} 
\centering 
\includegraphics[width=1\textwidth]{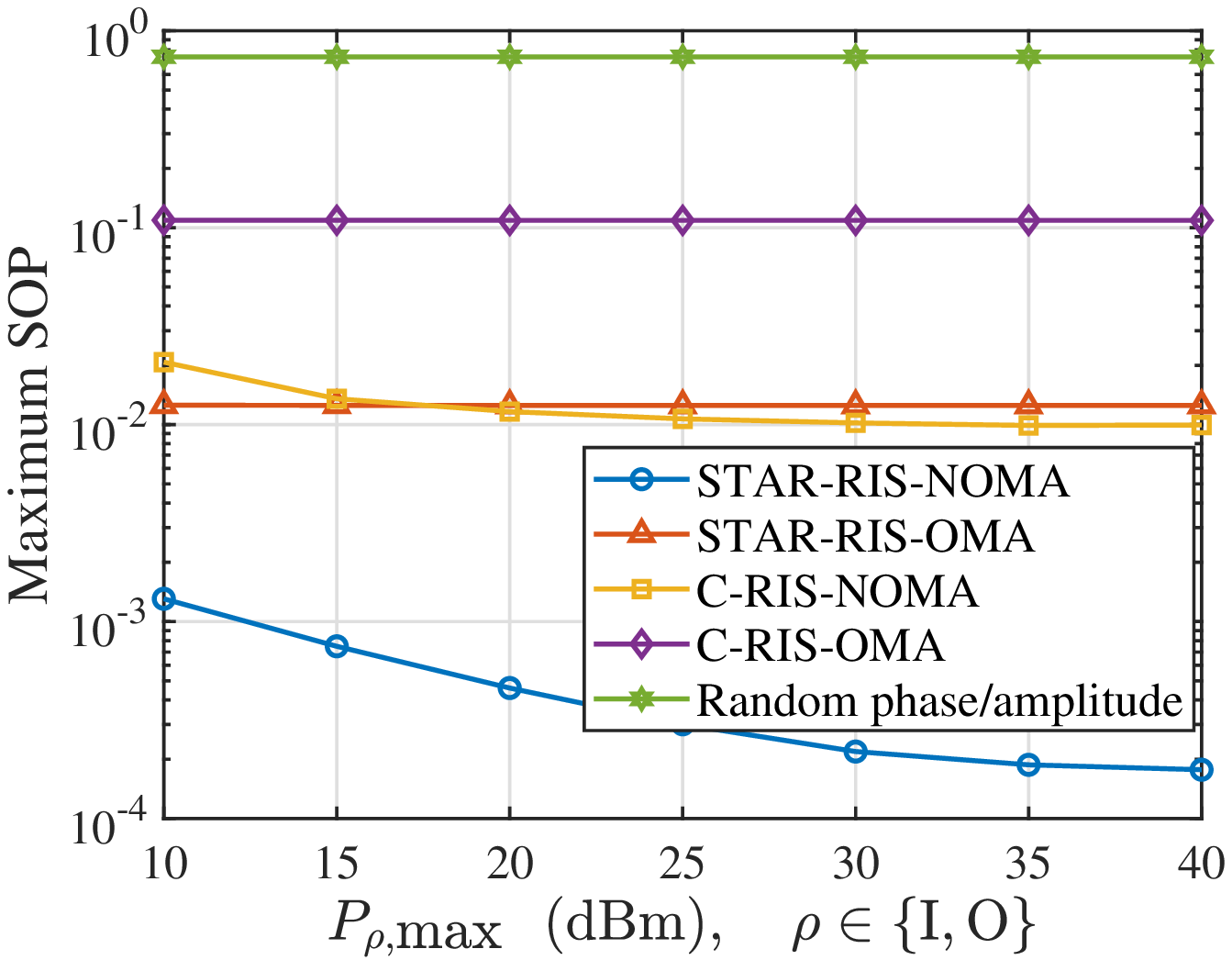} 
\caption{The maximum SOP among IU and OU versus the maximal transmit power of IU/OU with $N=20$, $M=8$, $R_{\text{c},\text{I}}=2$bps/Hz, $R_{\text{c},\text{O}}=0.5$bps/Hz, $R_{\text{s},\text{I}}=1.9$bps/Hz and $R_{\text{s},\text{O}}=0.4$bps/Hz.}
\label{Fig.7}
\end{minipage} \vspace{-8mm}
\end{figure}

Fig. \ref{Fig.6} and Fig. \ref{Fig.7} plot the minimum secrecy capacity and maximum SOP versus the maximal transmit power of IU and OU. To demonstrate the performance of the proposed scheme, we consider the following transmission schemes as the baseline schemes for comparison. \begin{itemize}
  \item \textbf{STAR-RIS-OMA}: For the STAR-RIS-OMA scheme, the BS serves IU and OU through the time division multiple access (TDMA) with the assistance of the STAR-RIS.
  \item \textbf{C-RIS-NOMA}: For the conventional RIS assisted NOMA (C-RIS-NOMA) scheme, we deploy a reflecting-only RIS and a transmitting-only RIS at the same location of the STAR-RIS to serve the IU and OU via the uplink NOMA principle, where each reflecting/transmitting-only RIS is equipped with $\frac{N}{2}$ elements for fairness comparison.
  \item \textbf{C-RIS-OMA}: For the conventional RIS assisted OMA (C-RIS-OMA) scheme, the BS serves IU and OU via the TDMA with the assistance of a reflecting-only RIS and a transmitting-only RIS.
  \item \textbf{Random phase/amplitude}: For the random phase/amplitude scheme, we consider to randomly generate the reflection/transmission phase shifts and amplitudes with satisfying $\theta^{\text{t}}_{n}, \theta^{\text{r}}_{n}\in(0,2\pi]$ and $\beta^{\text{t}}_{n}+\beta^{\text{r}}_{n}\leq 1$.
\end{itemize}

From Fig. \ref{Fig.6} and Fig. \ref{Fig.7}, we can observe that: 1) the STAR-RIS always provides a higher secrecy performance over the conventional RIS; 2) under the same RIS setting, the NOMA transmission outperforms the OMA transmission; and 3) the random phase/amplitude protocol suffers the worst performance in all transmission protocols. The reasons can be explained as follows. Since STAR-RIS can reflect and transmit the signals simultaneously, it is capable of adjusting the legitimate/eavesdropping channels more flexibly via the extra DoF. Also, unlike the conventional RIS employing half number of elements for wireless channel control, STAR-RIS employs all the elements to adaptively balance the legitimate communications and eavesdropping, which further enhances the secrecy performance of the network. On the other hand, NOMA protocol allows IU and OU to transmit their signals with the same time-frequency resource block, thus achieving better performance than OMA. Finally, as the random phase/amplitude protocol cannot unleash the passive gain brought by STAR-RIS, it behaves worst among all the schemes.

\begin{figure}[t]
\centering 
\begin{minipage}[b]{0.45\textwidth} 
\centering 
\includegraphics[width=1\textwidth]{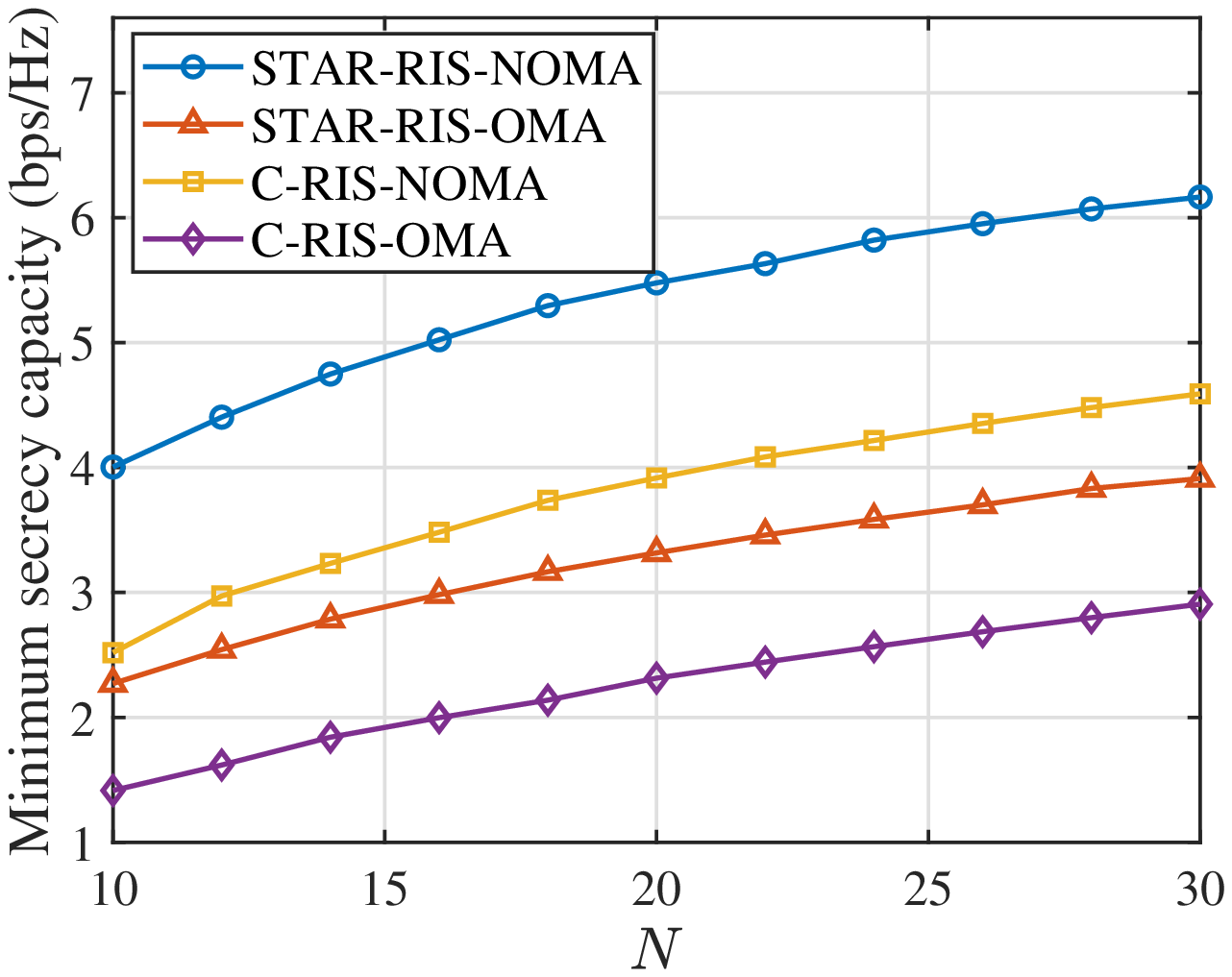}
\caption{The minimum secrecy capacity among IU and OU versus the number of relfection/transmission elements with $M=4$ and $P_{\text{I},\text{max}}=P_{\text{O},\text{max}}=15$dBm.}
\label{Fig.8}
\end{minipage}\qquad
\begin{minipage}[b]{0.45\textwidth} 
\centering 
\includegraphics[width=1\textwidth]{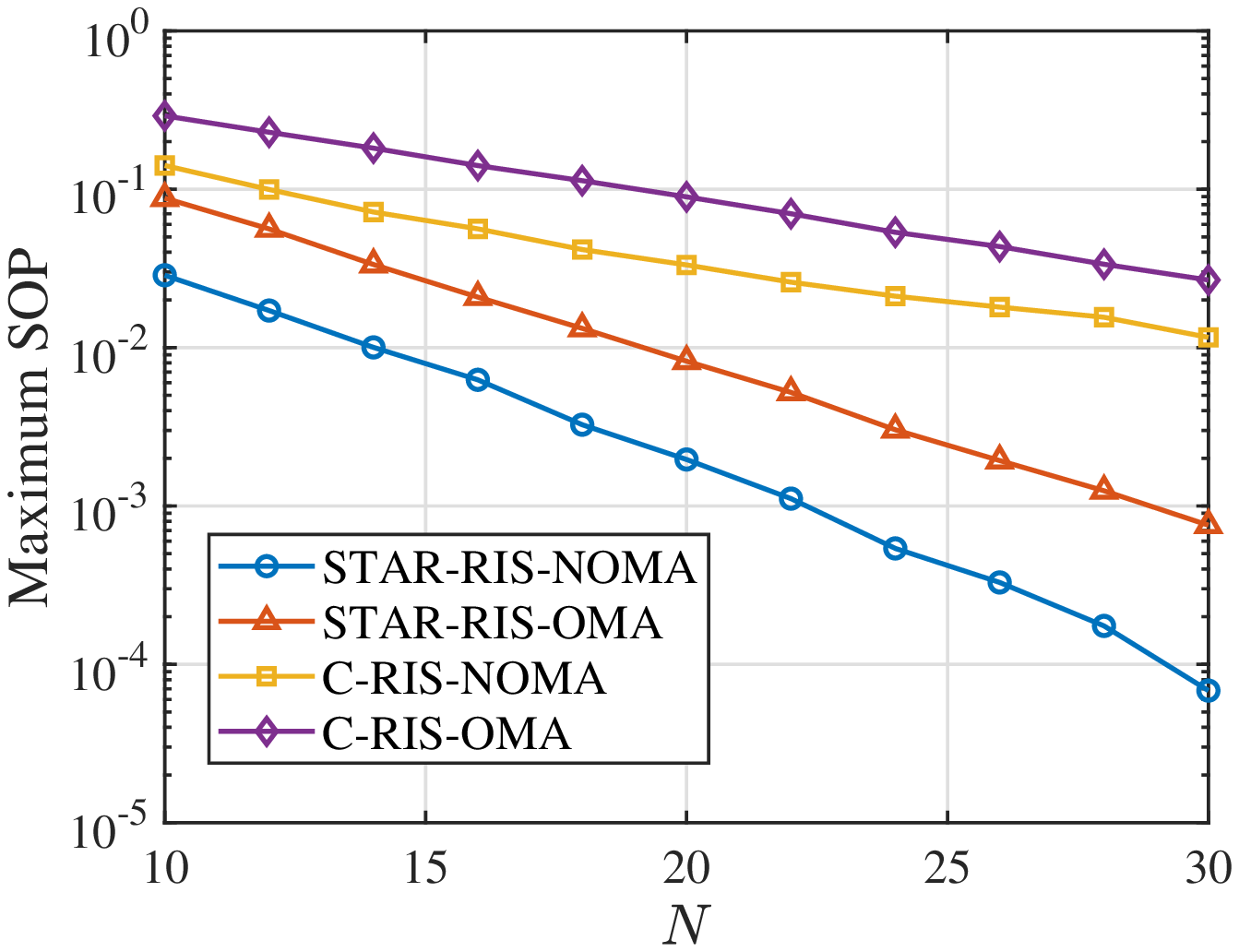} 
\caption{The maximum SOP among IU and OU versus the number of relfection/transmission elements with $M=4$, $P_{\text{I},\text{max}}=P_{\text{O},\text{max}}=15$dBm, $R_{\text{c},\text{I}}=2$bps/Hz, $R_{\text{c},\text{O}}=0.5$bps/Hz, $R_{\text{s},\text{I}}=1.8$bps/Hz and $R_{\text{s},\text{O}}=0.2$bps/Hz.}
\label{Fig.9}
\end{minipage} \vspace{-8mm}
\end{figure}

Fig. \ref{Fig.8} and Fig. \ref{Fig.9} illustrate the minimum secrecy capacity and maximum SOP versus the number of the reflection/transmission elements of the STAR-RIS. It can be observed that the secrecy performance achieved by the proposed scheme increases with the increase of $N$, which is because more reflection/transmission elements can provide larger passive beamforming design space for secrecy enhancement. Also can be observed, compared with C-RIS-NOMA, STAR-RIS-OMA achieves the higher secrecy performance in the case of full eavesdropping CSI, but behaves worse in the case of statistical eavesdropping CSI. This is due to the fact that in the case of full eavesdropping CSI, STAR-RIS/C-RIS is able to accurately adjust the eavesdropping channels, and thus, the power-domain multiplexing gains of NOMA take a leading role in the network secrecy enhancement. While in the case of statistical eavesdropping CSI, the transmit power needs to resist the eavesdropping channel uncertainty, so the secrecy performance of the network is dominated by the spatial gains brought by RIS, while STAR-RIS can provide more gains than conventional RIS.

\begin{figure}[t]
\centering 
\begin{minipage}[t]{0.45\textwidth} 
\centering 
\includegraphics[width=1\textwidth]{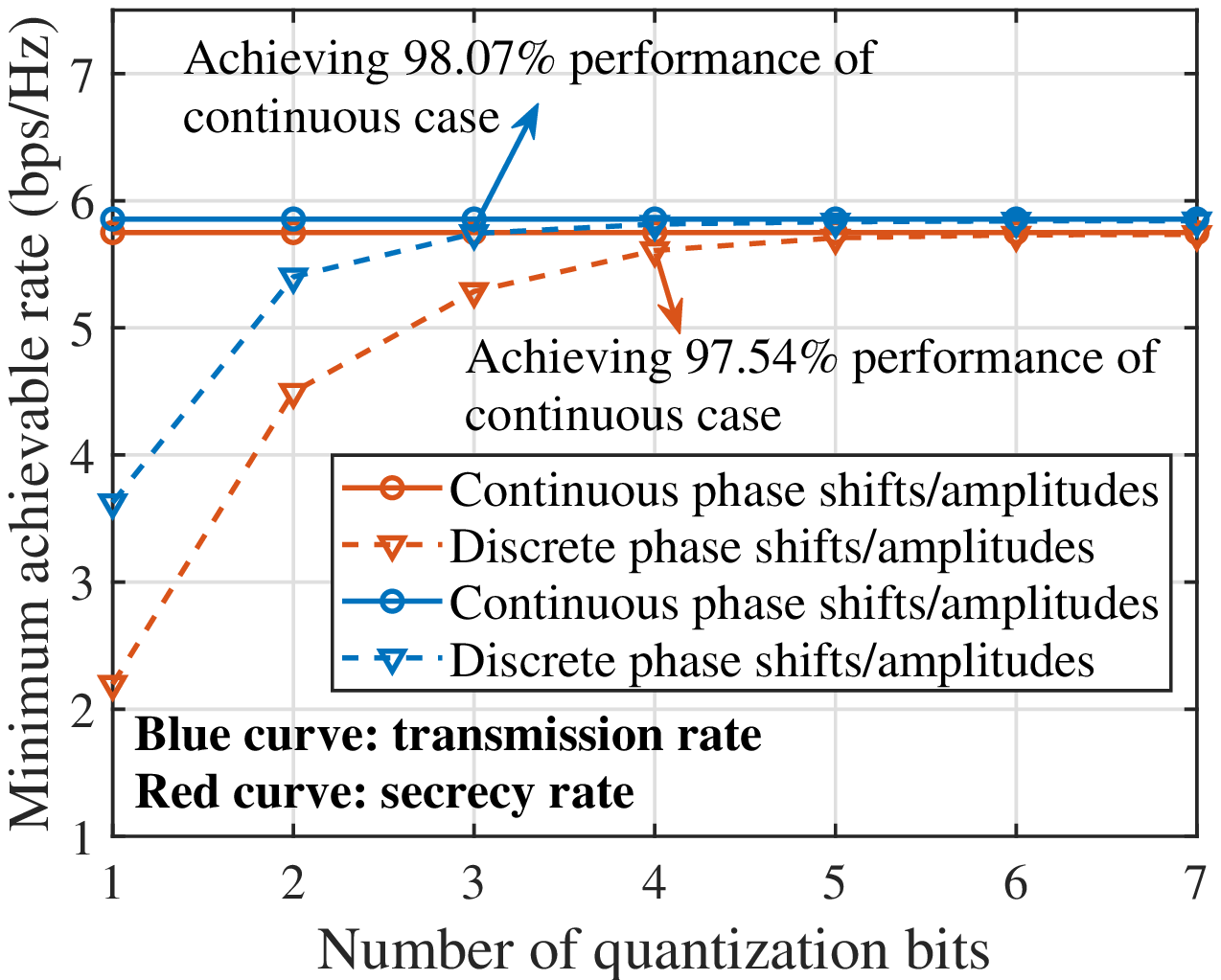}
\caption{The minimum achievable rate among IU and OU versus the number of phase shift and amplitude quantization bits with $M=8$ and $P_{\text{I},\text{max}}=P_{\text{O},\text{max}}=15$dBm.}
\label{Fig.10}
\end{minipage}\qquad
\begin{minipage}[t]{0.45\textwidth} 
\centering 
\includegraphics[width=1\textwidth]{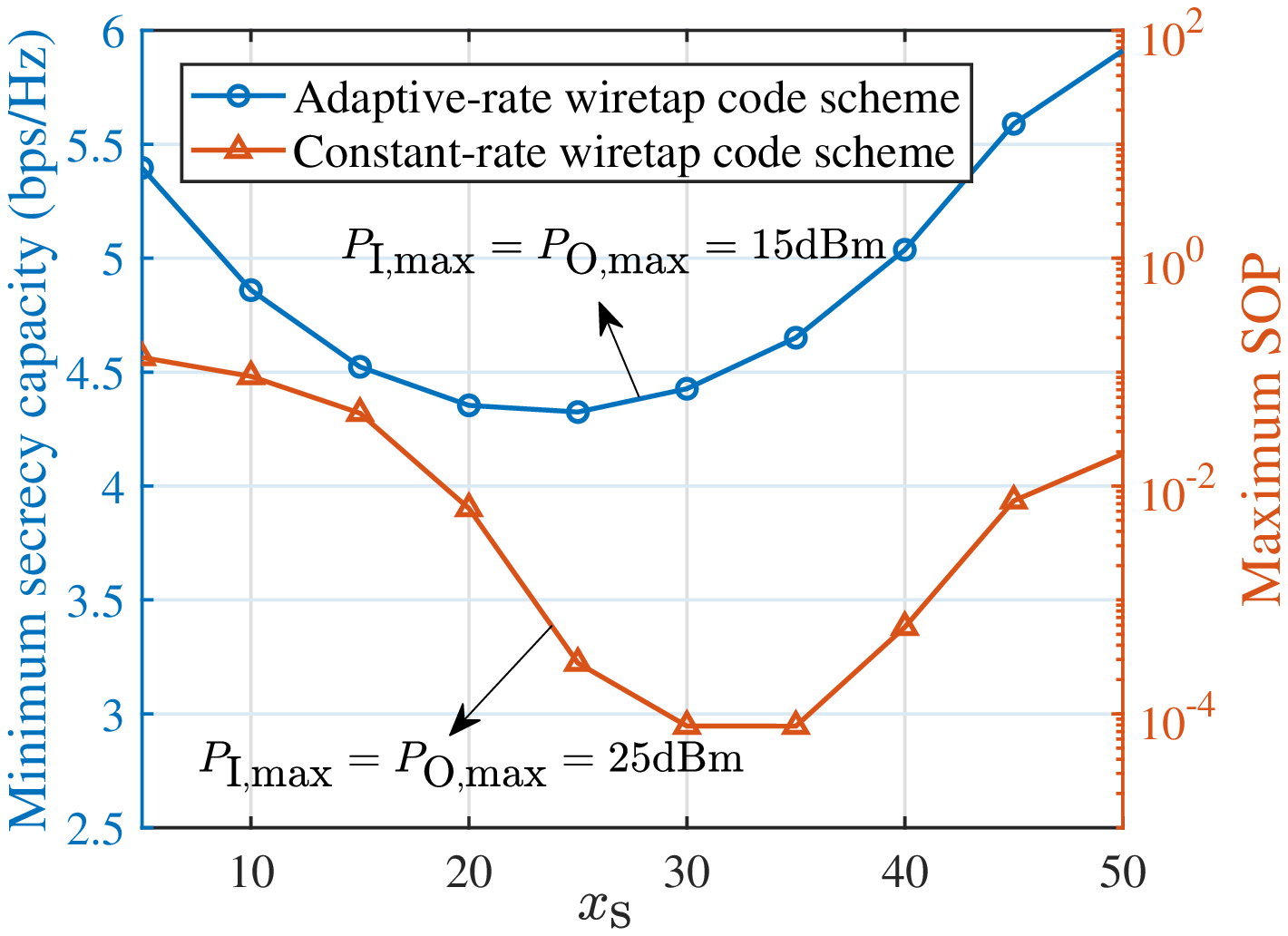} 
\caption{The minimum secrecy capacity and maximum SOP among IU and OU versus the location of STAR-RIS with $M=8$, $R_{\text{c},\text{I}}=R_{\text{c},\text{O}}=1$bps/Hz and $R_{\text{s},\text{I}}=R_{\text{s},\text{O}}=0.9$bps/Hz.}
\label{Fig.11}
\end{minipage} \vspace{-8mm}
\end{figure}

Fig. \ref{Fig.10} plots the minimum transmission rate and secrecy rate among IU and OU versus the number of the phase shift/amplitude quantization bits, where the transmission rate denotes the achievable rate without considering eavesdropper, while the quantized phase shift and amplitude are respectively determined as the values in the $q$-bit quantization sets $\{0,\frac{2\pi}{2^{q}},\dots,\frac{2\pi(2^{q}-1)}{2^{q}}\}$ and $\{0,\frac{1}{2^{q}-1},\frac{2}{2^{q}-1},\dots,1\}$, with being closest to the corresponding continuous values. It can be observed that even if two dimensions of quantization (i.e., phase shift and amplitude quantization) are considered in the STAR-RIS assisted transmissions, the transmission rate just requires $3$-bit quantization to achieve $98.07\%$ performance of the continuous phase shifts/amplitudes, while the $4$ bits quantized secrecy rate can only achieve $97.54\%$ performance of the continuous phase shifts/amplitudes. It is explained as 1) for the transmission rate, both phase shifts and amplitudes affect the analog beamforming design at the STAR-RIS, whereas the phase shifts adjustment plays a leading role; while 2) for secrecy rate, STAR-RIS needs to balance the legitimate signal strengthening and wiretap links suppressing, and thus, more accurate phase shifts/amplitudes control is required for secrecy guarantee.

In Fig. \ref{Fig.11}, we show the minimum secrecy capacity and maximum SOP among IU and OU versus the location of the STAR-RIS, where $x_{\text{s}}$ denotes the abscissa of the STAR-RIS. As can be observed, under the adaptive-rate wiretap code setting, it is better to deploy STAR-RIS near the users or BS for channel strengthening, while under the constant-rate wiretap code setting, it is preferable to deploy STAR-RIS far away from both of them for channel suppressing. It is since that: 1) for the adaptive-rate wiretap code scheme with full eavesdropping CSI, STAR-RIS is able to adjust the legitimate/eavesdropping channels exactly for controlling the codeword rate and redundant rate, which encourages the creation of strong cascaded links for unleashing the space potential of STAR-RIS; while 2) for the constant-rate wiretap code scheme with statistical eavesdropping CSI, STAR-RIS cannot effectively suppress eavesdropping and tends to focus on enhancing legitimate communications, which makes E more sensitive to the STAR-RIS location than BS. Meanwhile, due to the constant codeword rate setting, deploying STAR-RIS near the BS or users for channel strengthening brings no transmission rate increase to legitimate users. Hence, we should artificially deteriorate the cascaded links for the secrecy enhancement.

\section{Conclusion}

In this paper, a novel uplink transmission framework was proposed, in which STAR-RIS was employed to relay the superimposed signals from IU and OU to the BS while suppressing the malicious eavesdropping. Considering the different eavesdropping CSI assumptions, two joint beamforming optimization problems were respectively formulated for minimum secrecy capacity maximization and maximum SOP minimization. For the full eavesdropping CSI case, an AHB algorithm was proposed, where a polarization identity based two-layer iterative algorithm was developed to optimize the receive and passive beamforming, and the optimal transmit power policy was derived in the closed-form expression. As for the statistical eavesdropping CSI case, the exact SOP expression was derived and a modified AHB algorithm was further designed to minimize the maximum SOP among legitimate users. Numerical results demonstrated the secrecy superiority of the proposed scheme. It was also found that: 1) amplitude quantization had little impact on quantization performance, but the secrecy transmission significantly increased the required quantization resolution; and 2) \textmd{\textmd{\textmd{it}}} was better to deploy STAR-RIS around the users or the BS under the adaptive-rate wiretap code setting for cascaded channel enhancement, while it was preferable to deploy STAR-RIS far away from them for cascaded channel degradation under the constant-rate wiretap code setting.

\section*{Appendix A: Proof of Lemma 2}
The problem (24) is jointly convex for $\{\mathbf{U}_{\text{t}},\mathbf{U}_{\text{r}},\mathbf{W},\mathfrak{T}_{\rho,\text{t}/\text{r}}^{\text{lower}},
\mathfrak{T}_{\rho,\text{t}/\text{r}}^{\text{upper}},\varphi,\varrho_{\text{t}/\text{r}},\xi\}$ and meets Slater's constraint qualification \cite{Convex}. Therefore, the strong duality holds and the Lagrangian function with respect to $\mathbf{W}$ can be expressed as
    \begin{align}
    \label{A-1}\nonumber
    &\mathcal{L} = \lambda\text{Tr}(\mathbf{W})\!+\!\!\!\sum_{\rho\in\{\text{I},\text{O}\}}\!\mu_{\rho,1}\!
    \left(\|\mathbf{q}_{\rho}^{H}\mathbf{W}\mathbf{q}_{\rho}-\mathbf{U}_{\text{t}/\text{r}}\|_{F}^{2}\!-\!
    2\text{Tr}\left((\mathbf{q}_{\rho}^{H}\mathbf{W}\mathbf{q}_{\rho}+\mathbf{U}_{\text{t}/\text{r}})
        (\mathbf{q}_{\rho}^{H}\mathbf{\widetilde{W}}\mathbf{q}_{\rho}+\mathbf{\widetilde{U}}_{\text{t}/\text{r}})^{H}\right)\right)+\\ \tag{A-1}
    &\!\!\!\sum_{\rho\in\{\text{I},\text{O}\}}\!\!\!\mu_{\rho,2}\!
    \left(\|\mathbf{q}_{\rho}^{H}\mathbf{W}\mathbf{q}_{\rho}\!+\!\mathbf{U}_{\text{t}/\text{r}}\|_{F}^{2}\!-\!
    2\text{Tr}\left(\!(\mathbf{q}_{\rho}^{H}\mathbf{W}\mathbf{q}_{\rho}\!-\!\mathbf{U}_{\text{t}/\text{r}})
        (\mathbf{q}_{\rho}^{H}\mathbf{\widetilde{W}}\mathbf{q}_{\rho}\!-\!\mathbf{\widetilde{U}}_{\text{t}/\text{r}})^{H}
        \right)\!\right)\!-\!\text{Tr}(\mathbf{Y}\mathbf{W})\!+\!\tau,
    \end{align}
    where $\tau$ denotes the collection of the terms with no dependence of $\mathbf{W}$, while $\lambda\geq 0$, $\mu_{\rho,1}\geq 0$,  $\mu_{\rho,2}\geq 0$ and $\mathbf{Y}\succeq \mathbf{0}$ represent the Lagrange multipliers correspondingly. Checking the Karush-Kuhn-Tucker (KKT) conditions of $\mathbf{W}$, we list the structure of the optimal $\mathbf{W}^{*}$ as follows
    \begin{subequations}
    \begin{gather}
    \label{A-2a}\tag{A-2a} \text{K1}:\quad \lambda^{*}\geq 0, \quad \mu_{\rho,1}^{*}\geq 0,\quad \mu_{\rho,2}^{*}\geq 0, \quad \mathbf{Y}^{*}\succeq \mathbf{0},\\
    \label{A-2b}\tag{A-2b} \text{K2}:\quad \mathbf{Y}^{*}\mathbf{W}^{*}=\mathbf{0},\\
    \label{A-2c}\tag{A-2c} \text{K3}:\quad \nabla_{\mathbf{W}^{*}}\mathcal{L}=0,
    \end{gather}
    \end{subequations}
    where $\lambda^{*}$, $\mu_{\rho,1}^{*}$, $\mu_{\rho,2}^{*}$ and $\mathbf{Y}^{*}$ are the optimal Lagrange multipliers. The KKT condition K3 can be rewritten as
    \begin{equation}
    \label{A-3}\tag{A-3} \mathbf{Y}^{*} = \lambda\mathbf{I}-\bm{\Lambda},
    \end{equation}
    where $\bm{\Lambda}=\sum_{\rho\in\{\text{I},\text{O}\}}\mu_{\rho,1}(
    2(\mathbf{q}_{\rho}^{H}\mathbf{\widetilde{W}}\mathbf{q}_{\rho}+\mathbf{\widetilde{U}}_{\text{t}/\text{r}})^{H}+
    \mathbf{q}_{\rho}\mathbf{U}_{\text{t}/\text{r}}^{H}\mathbf{q}_{\rho}^{H}-
    \mathbf{q}_{\rho}\mathbf{q}_{\rho}^{H}\mathbf{\overline{W}}^{*}\mathbf{q}_{\rho}\mathbf{q}_{\rho}^{H})+
    \sum_{\rho\in\{\text{I},\text{O}\}}\mu_{\rho,2}\\(
    2(\mathbf{q}_{\rho}^{H}\mathbf{\widetilde{W}}\mathbf{q}_{\rho}\!-\!\mathbf{\widetilde{U}}_{\text{t}/\text{r}})^{H}-
    \mathbf{q}_{\rho}\mathbf{q}_{\rho}^{H}\mathbf{\overline{W}}^{*}
    \mathbf{q}_{\rho}\mathbf{q}_{\rho}^{H}-\mathbf{q}_{\rho}\mathbf{U}_{\text{t}/\text{r}}^{H}\mathbf{q}_{\rho}^{H})$. Here, we show that $\text{rank}(\mathbf{W}^{*})=1$ holds. First, to avoid an unbounded feasible solution space of the dual problem, $\lambda$ should be positive. With the KKT condition of $\mathbf{Y}^{*}\succeq \mathbf{0}$, $\rho_{\text{max}}(\bm{\Lambda})$ cannot be larger than $\lambda$, i.e., $\rho_{\text{max}}(\bm{\Lambda})\leq \lambda$. However, if $\rho_{\text{max}}(\bm{\Lambda})<\lambda$, we have $\lambda\mathbf{I}-\bm{\Lambda}\succ\mathbf{0}$, i.e., $\mathbf{Y}^{*}$ will be a full rank positive definite matrix. In this case, $\mathbf{W}^{*}=\mathbf{0}$ as KKT condition K2 insures that $\text{rank}(\mathbf{Y}^{*})+\text{rank}(\mathbf{W}^{*}) = M$, which however is not the optimal solution. Thus, we know that $\rho_{\text{max}}(\bm{\Lambda})=\lambda$ holds, which implies that $\mathbf{Y}^{*}$ is a positive semidefinite matrix with $M>\text{rank}(\mathbf{Y}^{*})\geq M-1$ \cite{D.Xu_rank-one}. In this case, we are able to span the null space of $\mathbf{Y}_{i}^{*}$ by an unit-norm vector $\mathbf{w}\in\mathbb{C}^{M\times 1}$, i.e., $\mathbf{Y}^{*}\mathbf{w}=\mathbf{0}$. Reviewing the KKT condition K2, $\mathbf{W}^{*}$ also lies in the null space of $\mathbf{Y}^{*}$. Thus, the optimal receiving beamforming matrix can be expressed as
    \begin{equation}
    \label{A-4}\tag{A-4}\mathbf{W}^{*} = \mathbf{w}\mathbf{w}^{H},
    \end{equation}
    which completes the proof.

\section*{Appendix B: Proof of Proposition 2}

In step $3$ of \textbf{Algorithm-2} at the $\textit{n}$th iteration, the stationary point solutions $\{\mathbf{W}^{*}(n), \mathbf{U}_{\text{t}/\text{r}}^{*}(n)\}$ are obtained via solving problem (24), which satisfy
\begin{equation}
    \label{B-1}\tag{B-1}R_{\text{s},\text{min}}(\mathbf{W}(n-1),\mathbf{U}_{\text{t}/\text{r}}(n-1),P_{\text{I}}(n),P_{\text{O}}(n))\leq
R_{\text{s},\text{min}}(\mathbf{W}^{*}(n),\mathbf{U}_{\text{t}/\text{r}}^{*}(n),P_{\text{I}}(n),P_{\text{O}}(n)).
\end{equation}
While in step $5$ of \textbf{Algorithm-2} at the $\textit{n}$th iteration, the optimal solutions $\{P_{\text{I}}^{*}(n),P_{\text{O}}^{*}(n)\}$ are obtained according to \eqref{31}, which meet
\begin{equation}
    \label{B-2}\tag{B-2}R_{\text{s},\text{min}}(\mathbf{W}(n),\mathbf{U}_{\text{t}/\text{r}}(n),P_{\text{I}}(n),P_{\text{O}}(n))\leq
R_{\text{s},\text{min}}(\mathbf{W}(n), \mathbf{U}_{\text{t}/\text{r}}(n),P_{\text{I}}^{*}(n+1),P_{\text{O}}^{*}(n+1)).
\end{equation}
Combining the above inequalities, it can be readily obtained that
\begin{equation}
    \label{B-3}\tag{B-3}R_{\text{s},\text{min}}(\mathbf{W}(n-1),\mathbf{U}_{\text{t}/\text{r}}(n-1),P_{\text{I}}(n),P_{\text{O}}(n))\leq
R_{\text{s},\text{min}}(\mathbf{W}(n), \mathbf{U}_{\text{t}/\text{r}}(n),P_{\text{I}}(n+1),P_{\text{O}}(n+1)).
\end{equation}
Meanwhile, $R_{\text{s},\text{min}}(\mathbf{W},\mathbf{U}_{\text{t}/\text{r}},P_{\text{I}},P_{\text{O}})$ is continuous over the compact feasible set of problem (8) \cite{Convex}, so the upper bound of the objective value is limited by a finite positive number, which proves the convergence of AHB algorithm. Moreover, since the optimization variables $\{\mathbf{W},\mathbf{U}_{\text{t}/\text{r}}\}$ and $\{P_{\text{I}},P_{\text{O}}\}$ restrict and influence each other in the alternating iterations, the converged solutions are usually some suboptimal solutions \cite{AO_conv}.

\end{document}